\documentclass[twocolumn]{emulateapj}
\usepackage{epstopdf}
\usepackage{graphicx}
\newcommand{\be}{\begin{equation}}
\newcommand{\ee}{\end{equation} }
\newcommand{\ba}{\begin{eqnarray}}
\newcommand{\ea}{\end{eqnarray}}

\newcommand{\bnabla}{\mbox{\boldmath$\nabla$}}

\newcommand{\nn}{\mbox{} \nonumber \\ \mbox{} }

\newcommand{\K}{{\rm K}}

\newcommand{\bxi}{\mbox{\boldmath$\xi$}}

\slugcomment{Submitted to the Astrophysical Journal}

\begin{document}  

\title{Global Crustal Dynamics of Magnetars in Relation to \\ Their Bright X-ray Outbursts}

\author{Christopher Thompson\altaffilmark{1}, Huan Yang\altaffilmark{2,3}, and N\'estor Ortiz\altaffilmark{2}}
\altaffiltext{1}{Canadian Institute for Theoretical Astrophysics, 60 St. George Street, Toronto, ON M5S 3H8, Canada}
\altaffiltext{2}{Perimeter Institute for Theoretical Physics, Waterloo, ON N2L 2Y5, Canada}
\altaffiltext{3}{Institute for Quantum Computing, University of Waterloo, Waterloo, ON N2L 3G1, Canada}

\begin{abstract}
This paper considers the yielding response of a neutron star crust to smooth, unbalanced Maxwell stresses imposed at
the core-crust boundary, and the coupling of the dynamic crust to the external magnetic field.  Stress buildup and
yielding in a magnetar crust is a global phenomenon:  an elastic distortion radiating from one plastically deforming
zone is shown to dramatically increase the creep rate in distant zones.  Runaway creep to dynamical rates is shown to
be possible, being enhanced by in situ heating and suppressed by thermal conduction and shearing of an embedded magnetic field.  A global and time-dependent model of elastic, plastic, magnetic, and thermal evolution is developed.  Fault-like
structures develop naturally, and a range of outburst timescales is observed.  Transient events with time
profiles similar to giant magnetar flares (millisecond rise, $\sim 0.1$ s duration, and decaying power-law
tails) result from runaway creep that starts in localized sub-km-sized patches and spreads across the crust.  
A one-dimensional model of stress relaxation in the vertically stratified crust shows that a modest increase in applied
stress allows embedded magnetic shear to escape the star over $\sim 3$-10 ms, dissipating greater energy if the 
exterior field is already sheared.  Several such zones coupled to each other naturally yield a burst of
duration $\sim 0.1$ s, as is observed over a wide range of burst energies.   The collective interaction of many plastic 
zones forces an overstability of global elastic modes of the crust, consistent with QPO activity extending over $\sim 100$ s.
Giant flares probably involve sudden meltdown in localized zones,
with high-frequency ($\gg 100$ Hz) QPOs corresponding to standing Alfv\'en waves within these zones.
\end{abstract}


\section{Introduction}\label{s:one}

Magnetars are neutron stars that emit intense and broad-band electromagnetic radiation by dissipating
very strong ($\gtrsim 10^{15}$ G) magnetic fields \citep{WT06,Mereghetti15,Turolla15}.  Our interest here is in
the extraordinary bursts of X-rays and gamma-rays that are detected sporadically from some magnetars.   We consider
how stress imbalances build up in the liquid core and crust of the neutron star, global effects of yielding in the crust,
and the coupling between crust and magnetosphere.

A role for the crust in mediating magnetar outbursts is suggested by i) the detection of quasi-periodic oscillations
(QPOs) during giant X-ray flares, many of which closely resemble global elastic modes of the crust 
\citep{israel05,sw05,ws06}; and ii) the similarity between the time for an elastic wave to propagate
around the star ($\sim 0.1$ s) and the duration of the most common Soft Gamma Repeater (SGR) bursts, as measured over a 
wide range of burst energies \citep{Gogus2000}.  SGR outbursts therefore appear to be a global phenomenon.   

Some magnetars -- the Anomalous X-ray Pulsars (AXPs) -- appear to dissipate their magnetic fields gradually, 
consistent with plastic creep within the active parts of their crusts \citep{TD1996,jones03}.  The SGRs behave similarly in
between their brief episodes of intense burst activity.  
Fluctuations in the X-ray output and spindown torque of magnetars show a broad range of timescales as well as of energies.
A central question is how changes in stress balance and current flow can emerge over timescales much shorter than the 
age of the star, but only rarely result in sub-second transient X-ray emission, and even then only in some sources.  

A promising approach starts with the extreme sensitivity of deformation rate in a solid to applied stress and temperature.
This is a principal focus of research in geodynamics \citep{turcotte02}, and the extension to the extreme densities
and pressures of neutron stars is supported by recent {\it ab initio} calculations using molecular dynamics methods
\citep{Chugunov2010,hoffman12}. 
Magnetar outbursts give evidence of very localized dissipation \citep{ibrahim01, lenters03, woods04, esposito07}.  
We show that narrow concentrations of rapid plastic flow are the natural response of the crustal solid to
relatively smooth, large-scale magnetic stresses acting on it from below.  We also investigate how magnetic
stresses are communicated from crust to magnetosphere on short timescales, resulting in intense X-ray emission.  Finally, we
argue that the global redistribution of stresses by elastic forces in the crust is tied to QPO activity during large flares, in other words, that global elastic modes may become overstable during large flares.

Burst-active and burst-quiet magnetars both persistently emit non-thermal X-rays with luminosity $\sim 10^{35}-10^{36}$ erg
s$^{-1}$, up to $\sim 10^3$ times what could be supplied by the spindown of the star.  This is consistent with the presence
of strong external electric currents (magnetospheric twist) in both bursting and non-bursting sources
\citep{TLK,BT07}.  The non-detection of energetically significant burst emission from several magnetars with active 
magnetospheres suggests to us that an external current-driven instability is not the primary driver of sub-second burst
activity, although such an instability may play a supporting role during episodes of fast and localized crustal shear deformation \citep{gill10,elenbaas16}.

\subsection{Overview of Results}

We begin by listing several open questions relating to magnetar activity, describe our approach to them, and the
conclusions we have drawn.  

The basic picture developed here views the magnetar crust as being composed of many interacting elastic 
units, which experience intermittent plastic creep over a wide range of rates.   Maxwell stresses imparted to the 
crust at its lower boundary source non-local solid stresses.  The creep rate in a given patch is determined
in part by the collective stress imparted by its neighbors, and is regulated by conductive cooling, neutrino emission,
and the shearing of an embedded magnetic field.  Localized zones of plastic failure result even if the applied
stresses are smooth.  In the most dynamic situations,
global crustal elastic modes become overstable when interacting with localized plastic patches.  

{\it Can the crust experience runaway deformation to dynamical rates, comparable to (shear wave speed)/(crustal scale height)?
Or are its deformations restricted to relatively slow plastic creep?}   We show that a slow increase in applied stress
within a small part of the crust, sourced by distant Maxwell stresses, can drive a very fast growth in creep rate.  In fact,
creep becomes fast enough that inertial forces start to counterbalance the Lorentz force and elastic force.  In this way,
the dense solid crust of a neutron star can momentarily behave like a fluid.

\begin{figure}
\epsscale{1.1}
\vskip -0.3in
\plotone{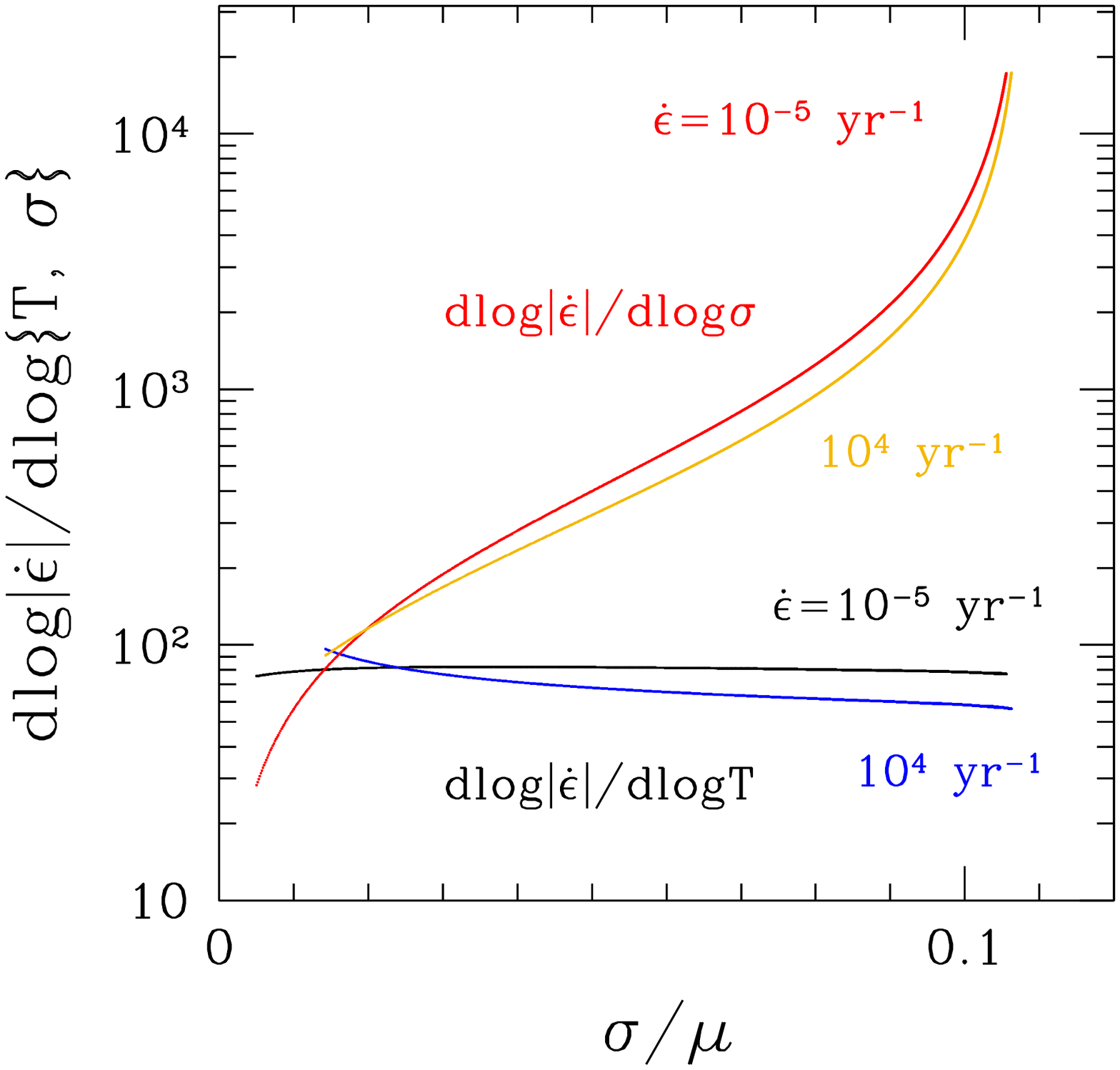}
\vskip -0.7in
\caption{Sensitivity of plastic creep rate in a Coulomb solid to changes in applied stress and temperature.   Creep
rate is determined by the expression (\ref{eqy1}) taken from \cite{Chugunov2010}.  Mass density $10^{14}$ g cm$^{-3}$, 
with nuclear composition taken from \cite{Negele1973}.  Range of stress plotted corresponds to temperature 
$0.99T_{\rm melt}$ (on the left) decreasing to $0.02\,T_{\rm melt}$ (on the right).  Creep rate is much more sensitive
to changes in stress than temperature, except when $T$ is close to the melt temperature $T_{\rm melt}$.}
\vskip .2in
\label{fig:dlogcreep}
\end{figure}

{\it What triggers a rapid growth in creep rate?}  Molecular dynamics simulations of dense Coulomb solids
\citep{Chugunov2010,hoffman12} show that the creep rate is much more sensitive to changes in stress, than it is to 
changes in temperature (Figure \ref{fig:dlogcreep}).

{\it What is the time profile of the dissipation?}  The net plastic dissipation rate in the crust peaks at a wide range of
rates, depending on the details of how stresses are configured.   In some cases, we find 
energy release profiles similar to those observed in giant magnetar flares:  very fast (millisecond) rise, 
$\sim 0.1$ s duration, followed by an extended power-law decay of plastic creep.

{\it How does fast creep couple to the magnetosphere?}  Magnetic twist or shear is directly ejected on 
a vertical shear wave timescale, consistent with some early theoretical ideas \citep{TD1995}.  
The efficiency of this process, as measured by the escaping 
Poynting flux, is greatly enhanced by the presence of a background magnetic twist extending into the magnetosphere.  
The energy that powers an SGR burst is {\it not}
primarily stored in an elastic wave, which only couples relatively slowly to magnetospheric modes \citep{Blaes1989, Link14}. 
The ejected magnetic shear may or may not oscillate in the magnetosphere, depending on the length of the excited field lines.

{\it What is the origin of the QPO activity seen in large magnetar flares?}
Stresses are redistributed throughout the magnetar crust by elastic forces.
Changes in the stress balance in one part of the crust will have a larger influence on a neighboring part 
than changes in temperature, both because of the greater sensitivity of creep rate, and because elastic
waves propagate much more rapidly than temperature fluctuations.  We find that modulation of the local creep rate
by a global elastic mode can feed back positively on the global mode.  

{\it How does a temperature increase influence subsequent yielding?}   The {\it duration and extent} of 
plastic creep is enhanced by in situ dissipation, which raises the temperature and reduces the equilibrium stress that 
can be sustained by a given patch of crust \citep{Chugunov2010}.   The Maxwell stresses imparted by the core
magnetic field are lowered gradually at distant points in the crust;  whereas there is a faster local effect on the 
$x$-$z$ and $y$-$z$ stress components (corresponding, e.g., to a redistribution of twist or shear along poloidal field lines).  For comparison, the thermal-magnetic front process proposed by
\cite{bl14} and \cite{li16} operates on an intermediate timescale.  It does not include a non-local contribution to
the stress, and therefore depends on very sharp gradients in temperature and Lorentz force to communicate changes in
strain rate. 

{\it What is the role of electron thermal conduction in regulating plastic creep?}   We find that thermal 
conduction generally reduces the creep rate, and in some circumstances can prevent runaway creep.  Equilibrium solutions for
a plastically deformation `spot' (0-dimensional) and `fault' (1-dimensional) can be constructed in which
the heat generated by plastic deformation is conducted away, or radiated to neutrinos.  

\begin{figure}
\plotone{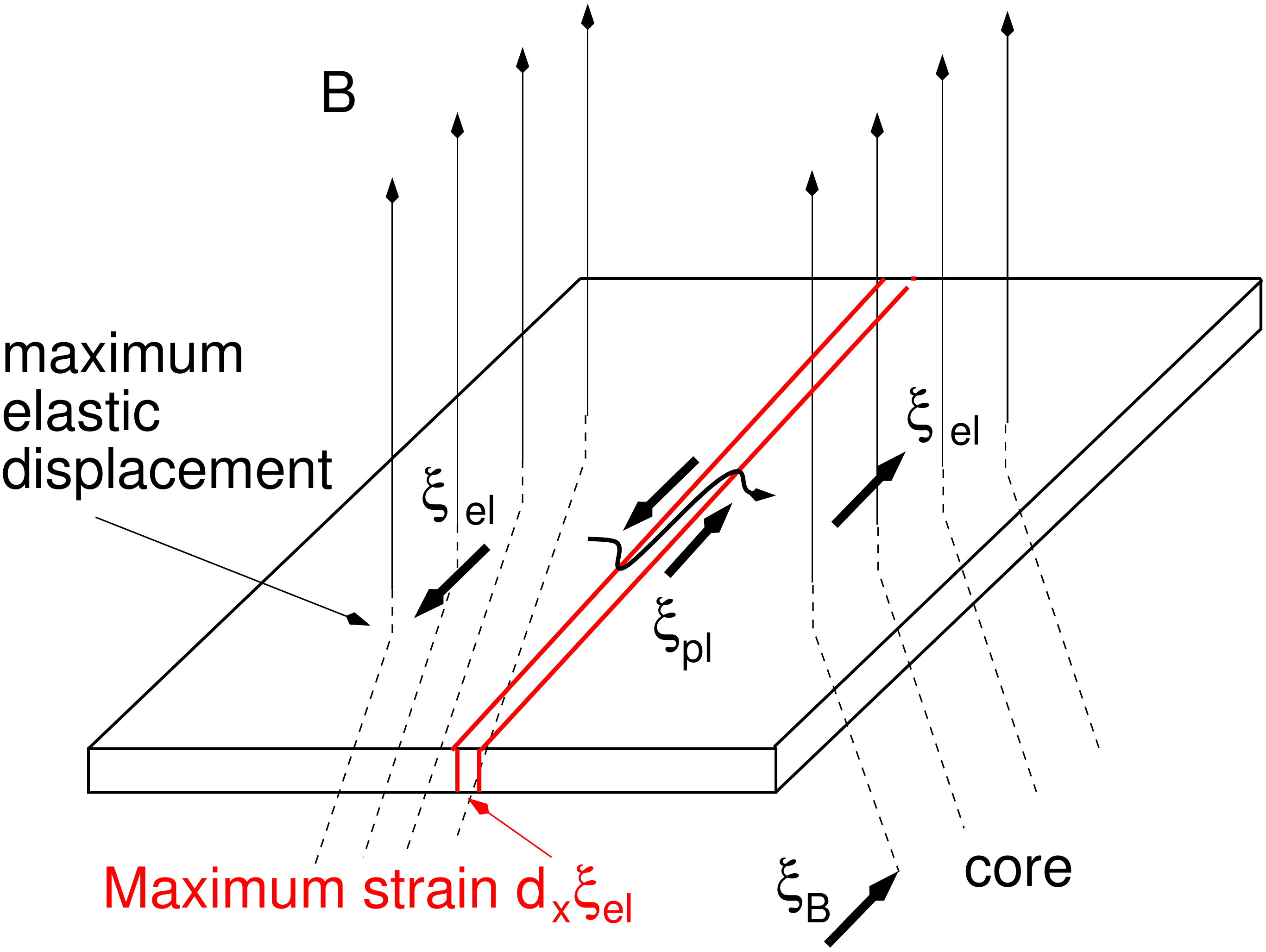}
\vskip .2in
\caption{An imbalance in Maxwell stress $B_\perp B_z/4\pi$ between the bottom and top of the magnetar crust
drives horizontal stresses that peak at distant locations.   The symmetric case shown here contains
a narrow zone of plastic creep, which bisects two zones of peak Maxwell stress and horizontal elastic displacement.
The problem of plastic creep in the crust is formulated here in terms of a displacement field $\bxi_B$ which 
imparts tilt to an initially vertical (radial) magnetic field.  A horizontal magnetic field embedded in the crust,
and oriented transverse to the plastic flow, plays an important role in compensating growth of the applied stress
and limiting creep.}
\vskip .2in
\label{fig:global_stress}
\end{figure}

{\it What is the relative importance of core and crustal magnetic fields on `breaking' the crust?}  Having fixed the magnitude
of the magnetic field, one obtains a larger solid stress from an imbalance in the $x$-$z$ and $y$-$z$ components
of the  Maxwell stress at the crust-core boundary,
as compared with the stress locally imparted by magnetic irregularities {\it within} the crust \citep{TD2001}.  The ratio of
the two is $R_\star/\delta R_c$, where $R_\star$ is the stellar radius and $\delta R_c$ the crustal thickness  
(Figure \ref{fig:global_stress}).
This Maxwell stress imbalance has a {\it non-local} effect, increasing the solid stress at distant points in the crust.
We find that the horizontal $x$-$y$ component of the Maxwell stress has the opposite effect, providing a significant 
buffer to runaway creep.

{\it How important is the role of hydromagnetic instability in driving the largest magnetar flares?}
Solid stresses in the crust can resist only a limited imbalance between hydrostatic, gravitational, and
elastic stresses.    We find that giant flare energies of $10^{45}$
erg or larger depend on the onset of a hydromagnetic instability in the magnetar core.  We also show that large
stress imbalances may develop spontaneously in the core on timescales as short as a day.

\begin{figure}
\vskip -0.3in
\epsscale{1.15}
\plotone{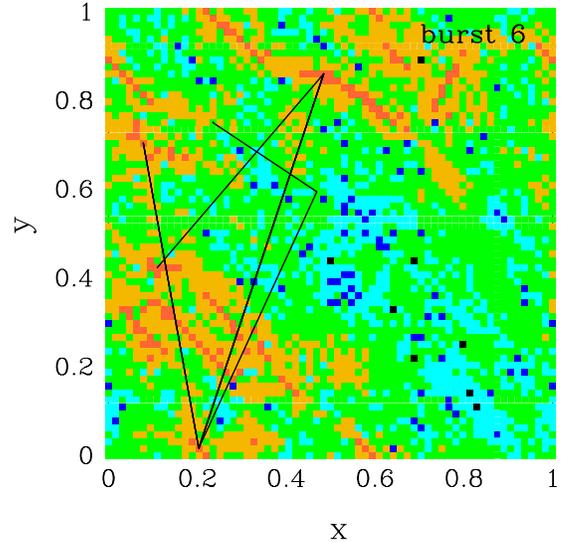}
\vskip -0.7in
\caption{Map of plastic dissipation in a planar solid with periodic boundary conditions.  The dissipation
has been integrated in time over a major outburst of total energy
$\simeq 1\times 10^{45}$ erg and peak duration $\sim 0.1$ s.  Black line shows the shifting location of
peak creep rate.   Red, gold, green, cyan, blue, 
pixels have energy release $> 10^{43}$ erg, $> 10^{42}$ erg, ..., $> 10^{39}$ erg.  The forcing 
Maxwell stress at the crust-core boundary has power spectrum $dB^2/d\ln k \propto (k_\parallel^2 + 0.3 k_\perp^2)^{-1/2}$,
where $k_\perp/k_\parallel$ are the components of the 2D wavevector perpendicular/parallel to the axis along which power
is concentrated.  This axis is chosen as a $45^\circ$ diagonal so as to suppress grid-alignment effects.}
\vskip .2in
\label{fig:2Danis}
\end{figure}

{\it Can fault-like features form in the magnetar crust?}   We find that if the crust is stressed from below in 
a slightly anisotropic manner (corresponding e.g. to a large-scale winding of the core magnetic field) then
concentrated zones of strong shear develop naturally (Figure \ref{fig:2Danis}).  
Adding a horizontal magnetic field to the crust does not suppress this effect.  The formation of open cracks 
is inhibited by the extreme ratio of hydrostatic pressure to shear modulus in the crust of a neutron star \citep{jones03}, 
Nonetheless, the small lengthscale $\delta R_c \sim 0.03/R_\star$ provides a small-scale cutoff for
global elastic stresses, and defines a characteristic fault width.  Our basic approach to calculating the
global evolution of horizontal solid stresses by dividing the crust into $\sim 4\pi (R_\star/\delta R_c)^2 \sim 10^4$ 
elastic units.  

{\it Why are all magnetars not bursting SGRs, and why are SGRs not bursting all the time, given the extreme
non-linearity and temperature sensitivity of the creep rate?}  We find that the Coulomb solid attains a 
state of equilibrium creep (at rates intermediate between the dynamical rate and the inverse age of the star)
only if stress growth is buffered by the shearing of an embedded magnetic field.   
As this transverse field weakens, the crust becomes more sensitive to slow changes in applied stress.  
A tangled magnetic field is most effective at limiting runaway 
large-scale creep.  Hall drift is a plausible source of small-scale magnetic irregularities \citep{GR92}, and
in this way may actually limit SGR burst activity.

\cite{LL12} studied the feedback of magnetic field stretching on crustal shear, suggesting that it could
suppress the formation of strong shear layers, thereby pointing to an external trigger for magnetar flares.  
We find that this suppression is not complete, due to the non-local nature of the force balance in the 
thin magnetar crust.  Nonetheless, field line stretching does appear to play a key role in desensitizing
the magnetar crust to changes in applied stress and temperature.

\subsection{Where is the Magnetic Field Anchored?}

In contrast with the approach taken here, a number of calculations of magnetic field evolution 
assume that the magnetic field is anchored entirely in the magnetar crust, setting aside any effects of a poloidal
field threading crust/core boundary (e.g. \citealt{perna11,bl14,gourg16}, and references therein).
This is done, in part, for reasons of computational facility:  one can then neglect the response of the
liquid core to changing Maxwell stresses.

It is therefore worth asking whether a purely crustal magnetic field is a realistic configuration in a neutron star.
Considerations of the origin of this field suggest that it is not.
The progenitor star experiences a series of hydrodynamic instabilities which successively modify the magnetic
field, starting with main sequence core convection and growing in power to the most extreme core collapse phase \citep{TD1993}.
Post-collapse accretion following the brief stalling of a bounce shock deposits some $3-10$ times
the mass of the eventual solid neutron star crust.  This allows a transient activation of the magnetorotational 
instability in a rarefied mantle \citep{AW03}, which is followed by more persistent linear winding of the magnetic
field as the mantle collapses.  Rapid neutrino heating drives buoyant motions of the toroidal magnetic field
if the seed field exceeds a critical value \citep{TM2001}, thereby potentially distinguishing magnetars from 
neutron stars with a combination of slower rotation and weaker seed fields.   

The dilution of the core magnetic field by entrainment with outward-drifting superfluid vortices \citep{ruderman98}
is impeded in comparison with ordinary radio pulsars by a relatively low ratio of rotational to magnetic energy.  
Given that the core superfluid transition takes place at age $\sim 300$ yr in a more weakly magnetized neutron star
such as the Cas A remnant \citep{Ho15}, one sees that a magnetar must reach a spin period $> 1 $ s before superfluid 
vortices appear in its core.  Heating of the cores of the most active magnetars can delay this transition further:
indeed an active lifetime of $\sim 10^3-10^4$ yr was used to infer a core superfluid transition temperature
$\lesssim 6\times 10^8 \K$ \citep{arras04}.

We therefore adopt the following simplified magnetic field configuration:  the field is strongly tilted or
twisted in the core, but becomes essentially vertical in the crust.  Of course a hybrid configuration is
plausibly attained in magnetars, with the axis of the magnetic field winding tilted with respect to the local
direction of gravity.  But the chosen configuration is adequate to demonstrate the effects we are interested in.

Empirically there appears to be a strong correlation between the detection of dense episodes of
short burst activity from magnetars, and their ability to generate a giant flare.  This suggests that
the same structures responsible for the giant flares also give rise to the shorter bursts.  The
giant flares, in turn, appear to depend on large-scale unbalanced stresses in the neutron star, which
partly motivates our approach that starts with a core magnetic field making a transition out of magnetostatic
equilibrium.

\subsection{Plan of the Paper}

The plan of this paper is as follows.  Section \ref{s:two} analyzes the plastic response of a small patch
of Coulomb solid to an externally imposed stress.  

Section \ref{s:three} sets up the problem of plastic flow and elastic response in two dimensions,
in response to an inhomogeneous Maxwell stress that is applied at the base of the crust.  

Section \ref{s:four} analyzes the results of global time-dependent calculations
of such a stressed model crust.  Section \ref{s:five} presents outburst
light curves; and shows how the energy dissipation is distributed between magnetic, plastic and elastic components,
and how the peak of dissipation circulates around the crust

In Section \ref{s:six} we turn to consider the vertical redistribution of stresses in a solid with a strong
density stratification, appropriate to a neutron star crust.  
Fast ejection of magnetic twist is encountered when a solid crust experiences runaway
creep in response to an externally applied stress.  This provides, we believe, a promising foundation for
understanding the short ($\sim 0.1$ s) SGR bursts.   

The slow damping of free vertical Alfv\'enic oscillations in liquified patches of the crust is described in 
Section \ref{s:seven}.  This suggests that the high-frequency QPOs
observed in magnetar outbursts \citep{sw06} may involve the vertical excitation of the magnetic
field in melted parts of the crust (e.g. within heated fault zones).   

The interaction between the local Maxwell stresses and global elastic oscillations is addressed in Section
\ref{s:eight}:  we show that the global mode is overstable if the localized stresses have a `patch' like
geometry.

The concluding Section \ref{s:nine} makes a comparison with independent theoretical approaches and summarizes
the implications of our work for the physics of magnetars.

\section{Stability of Plastic Flow \\ in Reduced Dimensions}\label{s:two}

We begin by considering the simplest problem of local plastic flow in response to 
an externally imposed stress.  The non-local origin of this stress within
a magnetically deformed crust is described in Section \ref{s:three}.

We include the effects of in situ heating, thermal conduction,
neutrino cooling, and the shearing out of a magnetic field flowing perpendicular to the
axis of creep.  This is done first in a 0-dimensional model, followed by planar 
(1-dimensional) geometry (e.g., a fault).  

We take the plastic deformation locally to have cartesian symmetry, e.g., to be described
by a scalar function $\dot\varepsilon_{\rm pl} = \partial_x v_y$.  The dependence of 
$\dot\varepsilon_{\rm pl}$ on solid stress $\sigma$ and temperature has been calibrated by
molecular dynamics calculations.  We adopt the following relation\footnote{Molecular dynamics 
of neutron star crustal material do not presently capture the effects of large-scale networks 
of defects on the stress-strain relation, e.g, effects such as strain hardening, but in the 
absence of direct measurements they provide the firmest basis for calculating the evolution 
of magnetar crusts.} from \cite{Chugunov2010},
\be\label{eqy1}
\frac{\dot{\varepsilon_{\rm pl}}}{\omega_p} = {1\over 0.183\,\bar{N}\Gamma}
\,e^{-\bar U \Gamma+\bar{\sigma} \bar{N} \Gamma}.
\ee
Here $n$ is the ion number density, $\omega_p$ is the ion plasma frequency, $e$ is the electron 
charge, $+Ze$ is the average charge per ion, $a$ is the mean separation between ions, $\Gamma = Z^2 e^2/(a T)$ is 
the melting parameter, $\bar{\sigma} = \sigma /(n Z^2 e^2/a)$, and  
\be
\bar{U} = 0.366,\, \bar{N} = \frac{500}{\Gamma-149}+18.5.
\ee

The melting temperature $T_m$ of the crystal corresponds to $\Gamma_m \simeq 176$ \citep{stringfellow90}.
Through out this paper, we adopt a low-temperature approximation for the shear modulus
$\mu \approx 0.183 \, n Z^2 e^2/a$ \citep{Strohmayer1991}.  Finite-temperature corrections
to $\mu$ are generally less important than the temperature dependence of the creep rate.
We also neglect the effects of non-spherical nuclei, which can weaken the crust \citep{sotani11}:
when considering vertically averaged stresses, we focus on a density $10^{14}$ g cm$^{-3}$.
Then Equation (\ref{eqy1}) can be re-written as
\be\label{eqy}
\frac{0.183 \sigma}{\mu} =\frac{\bar{U}}{\bar{N}}+\frac{1}{\bar{N} \Gamma} 
\log \left [0.183\frac{\dot{\varepsilon_{\rm pl}}}{\omega_p} \bar{N}\Gamma \right ].
\ee

Expression (\ref{eqy1}) contains the familiar Boltzmann factor $e^{-\bar U\Gamma}$ representing
thermal activation of dislocation drift within the solid.  The second term in the exponential
is more important for the {\it initiation} of fast creep, in the sense that the creep rate is
more sensitive to changes in stress than to changes in temperature (Figure \ref{fig:dlogcreep}). 
We typically consider a starting temperature well below the melt temperature
at the base of the crust.  Then $\Gamma$ starts as a large number where most of the magnetic shear energy is
concentrated, and the first term in the right-hand side of Equation (\ref{eqy}) dominates.  

One sometimes describes a yielding solid as a very viscous fluid with a temperature-dependent
yield stress -- see \cite{bl14} and \cite{li2015} for considerations of magnetar crusts -- but in
such an approach the strong exponential dependence of creep rate on applied stress is only roughly captured.

\begin{figure*}
\vskip -0.3in
\epsscale{1.15}
\plottwo{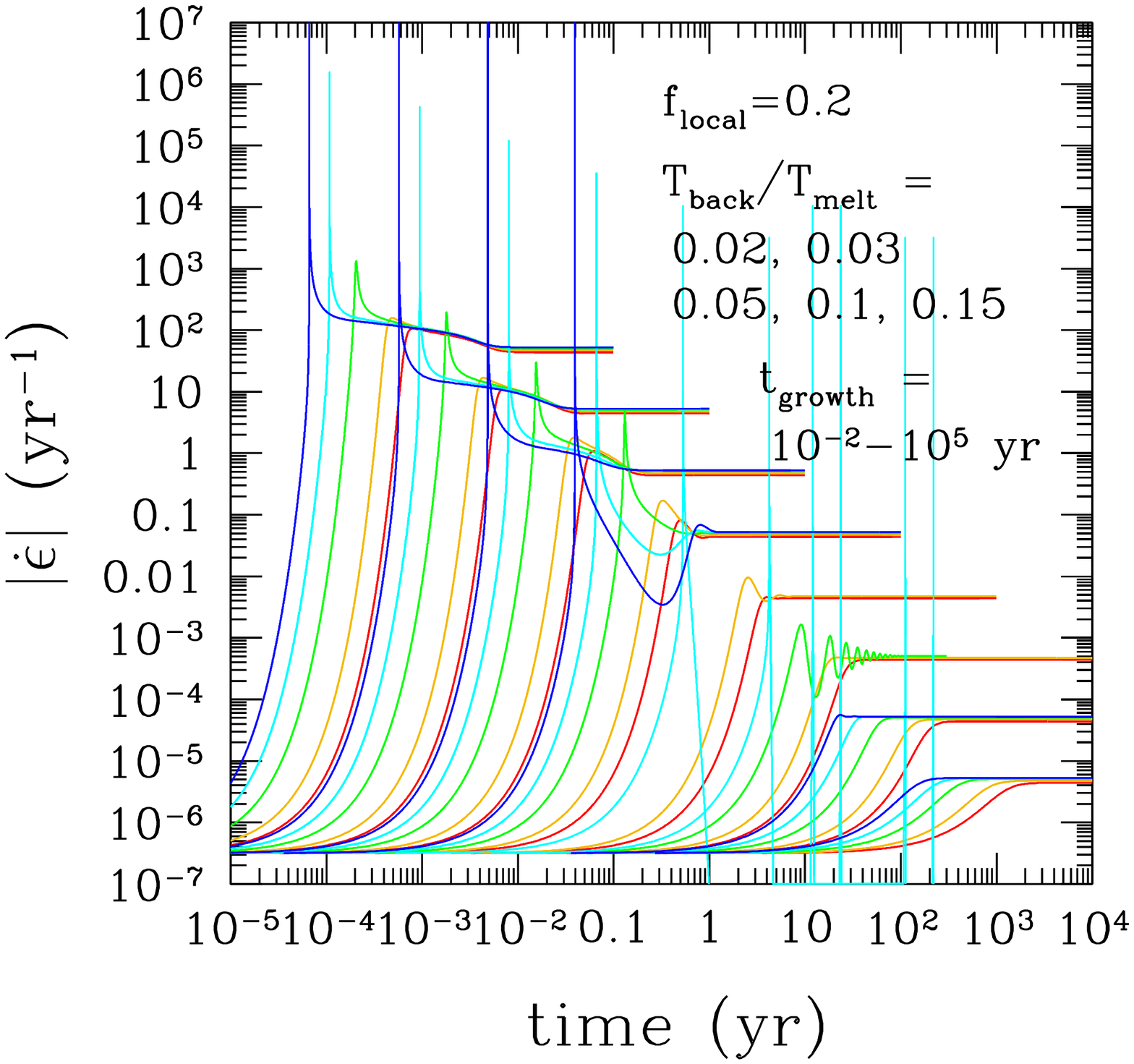}{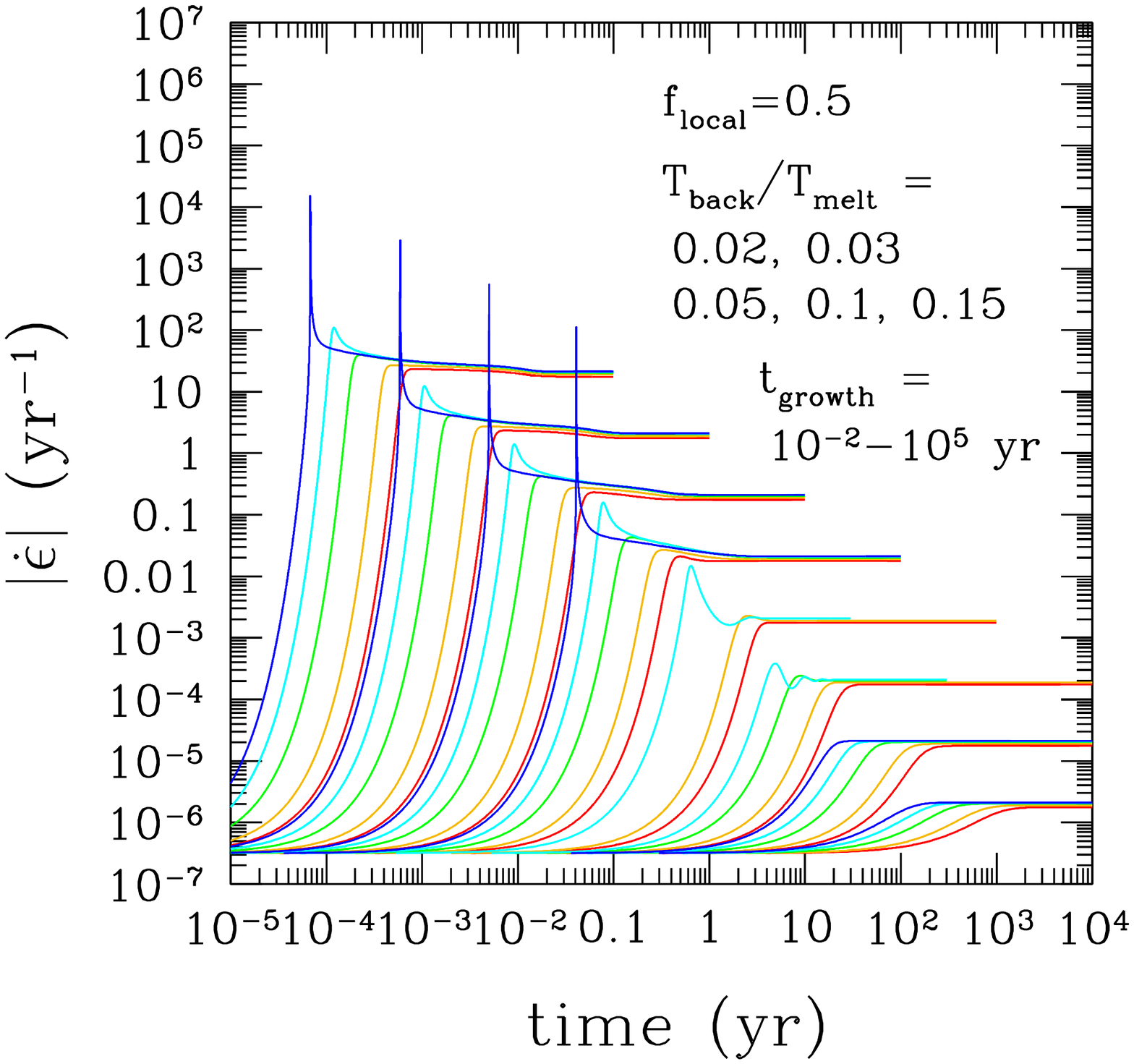}
\vskip -1in
\plottwo{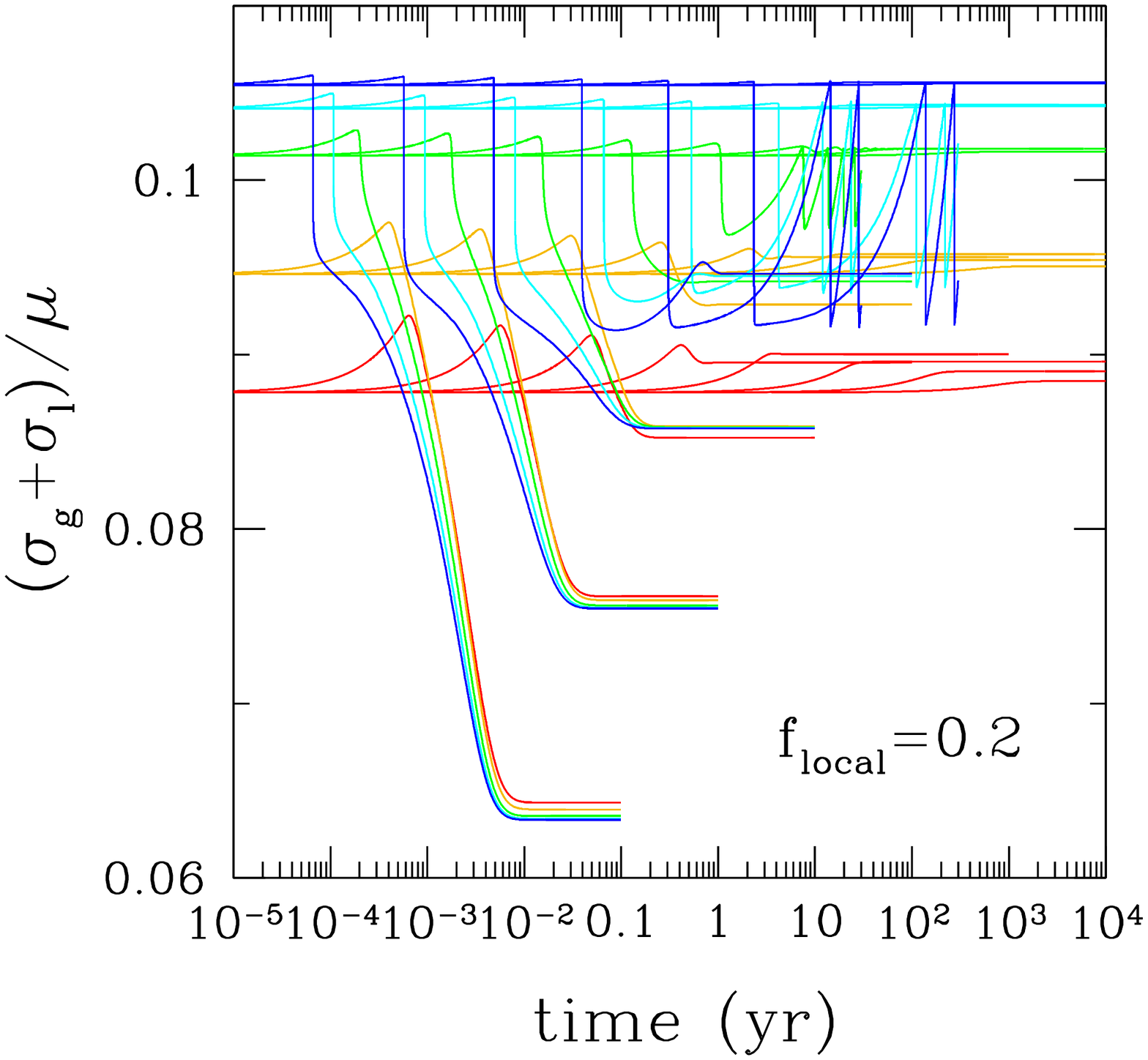}{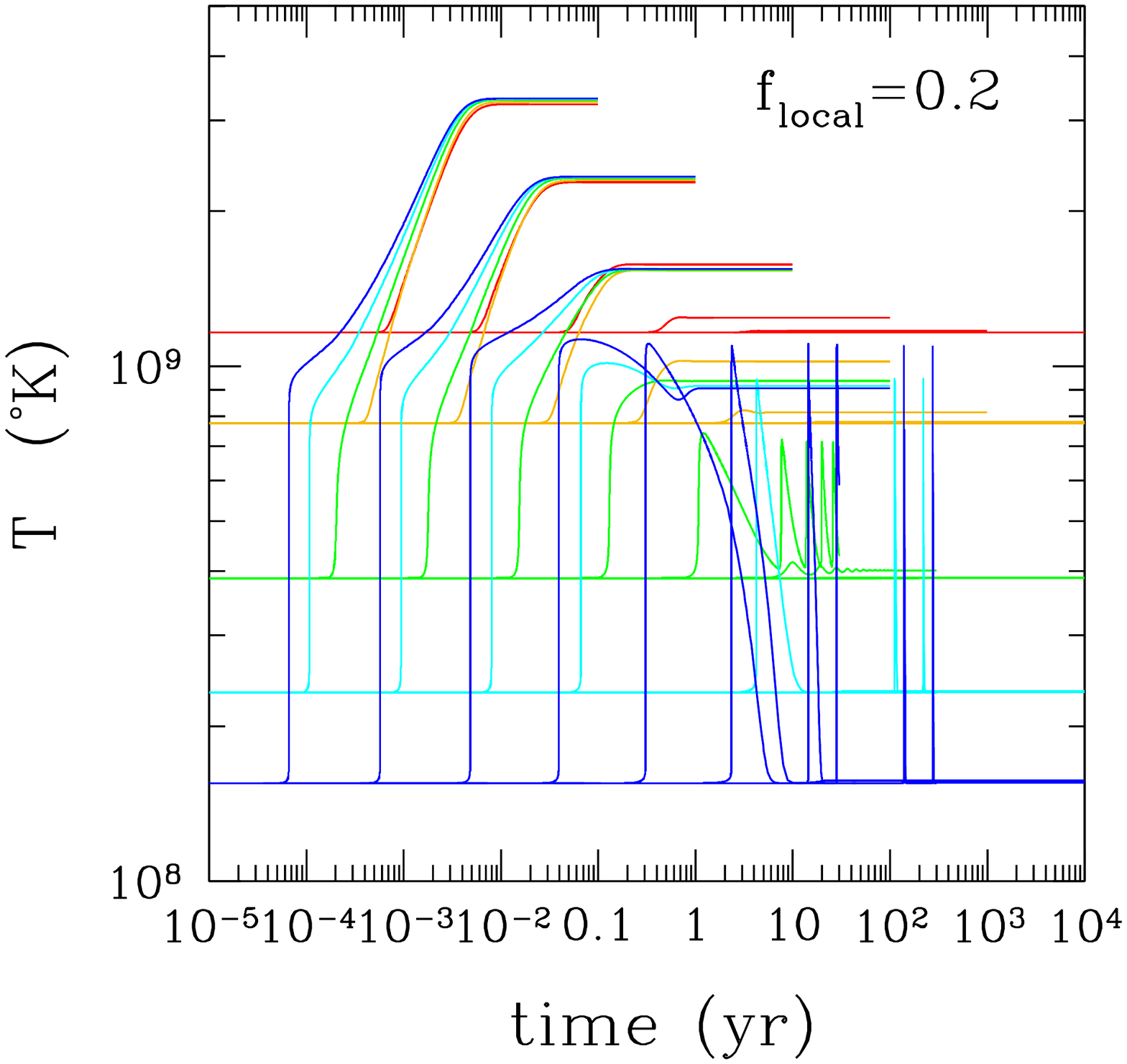}
\vskip -0.7in
\caption{{\it Top panels:}  Time dependence of creep rate in a small patch of magnetar crust that is exposed
to an external stress with growth time $10^{-2} - 10^5$ yr, uniformly spaced in ${\rm log}_{10}t_{\rm growth}$.
Curves with faster stress growth lie further to the left.  
Density $10^{14}$ g cm$^{-3}$ and background temperature $T_{\rm base}$ ranging from $2\times 10^8 \K$
to $6\times 10^9 \K$ (corresponding to 0.02-0.6 of the melt temperature, blue to red curves).
Oscillations appear when the growth time is comparable
to the conductive cooling time.   Plateaus in creep rate that are obtained after an initial spike correspond to
a steady state in which plastic heating is balanced by neutrino cooling.  {\it Bottom left panel}:  corresponding change
in the total stress.  {\it Bottom right panel:}  corresponding change in temperature.  There is a significant reduction in 
stress as temperature builds up.}
\vskip .2in
\label{fig:zeroD}
\end{figure*}

\subsection{Plastic Spot with Imposed Global Stress and \\ Compensating Local Maxwell Stress}

We consider the solid stress to be the sum of local and global contributions,
\be\label{eq:sig}
\sigma = \sigma_l + \sigma_g = \varepsilon_{\rm el,\it l}\,\mu  + \sigma_g.
\ee
The local stress $\sigma_l$ compensates a local Maxwell stress that grows in magnitude in response to 
plastic shear. Here $\perp$ and $\parallel$ label coordinates perpendicular and parallel to the direction of plastic
flow.  The applied global stress $\sigma_g$ is taken to grow on a timescale $t_{\rm growth}$.  It is
sourced by Maxwell stresses at a distance $L_x \gg \delta R_c$ from the plastic zone, and varies slowly
within the plastically deforming patch.  

We work in a frame where the plastic flow velocity $v_\parallel$ vanishes at a given point in the solid, so that
\be
v_\parallel = (\dot\varepsilon_{\rm pl} + \dot\varepsilon_{\rm el,\it l})x_\perp
\ee
in the direction transverse to the flow.  Here there is an additional contribution to the solid displacement 
from the elastic response to an imbalance between $\sigma_l$ and $B_\perp B_\parallel/4\pi$, which 
develops as the result of plastic creep.
The system moves through a series of equilibrium states in which $\partial_\perp (B_\perp B_\parallel/4\pi + 
\sigma_l + \sigma_g) = \partial_\perp(B_\perp B_\parallel/4\pi + \sigma_l) = 0$.  Hence,
\be
{B_\perp \partial_t B_\parallel\over 4\pi} + \partial_t\sigma_l = 
           {B_\perp \partial_t B_\parallel\over 4\pi} + \dot\varepsilon_{\rm el,\it l}\,\mu = 0,
\ee
where
\be\label{eq:Bpar}
\partial_t B_\parallel = \partial_\perp (v_\parallel B_\perp) = 
(\dot\varepsilon_{\rm pl} + \dot\varepsilon_{\rm el,\it l}) B_\perp\, .
\ee
Combining these equations, one sees the local stress changes in a negative manner with respect to the 
global stress, so as to partly compensate its growth, 
\ba\label{eq:locsig}
\partial_t\sigma_l &=& -{B_\perp^2/4\pi \mu\over 1 + B_\perp^2/4\pi\mu} \dot\varepsilon_{\rm pl}\,\mu
\equiv -f_{\rm local}\cdot \dot\varepsilon_{\rm pl}\,\mu;\nn
\partial_t\sigma_g &=& 
{\sigma_{g,0}\over t_{\rm growth}} -  (1-f_{\rm local}) {\dot\varepsilon_{\rm pl} \delta R_c\over L_x} \mu.
\ea
The equation for $\sigma_g$ includes a second term representing the decrease in large-scale elastic stress
caused by the net displacement $v_\parallel \sim (\dot\varepsilon_{\rm pl} + \dot\varepsilon_{\rm el,\it l}) \delta R_c
= (1-f_{\rm local})\dot\varepsilon_{\rm pl}\delta R_c$ across the plastic patch.

The energy equation is evolved at density $10^{14}$ g cm$^{-3}$, where the global stress is concentrated 
(Section \ref{s:three}),
\be\label{eq:en}
C_V\partial_tT = \dot\varepsilon_{\rm pl}(\sigma,T) \sigma - \ell^{-2} K_{\rm cond} (T-T_{\rm base}) - \dot Q_\nu(T,B).
\ee
In Equation (\ref{eq:en}), $C_V$ is the specific heat, which is approximated by the contribution of the ion lattice
\citep{chabrier93} with the effective ion mass raised by the entrainment of a fraction $0.3$ of superfluid neutrons 
\citep{onsi08}.  Then $C_V \sim T^3$ at
temperatures below the Debye temperature, meaning that the creep rate is more sensitive to heat input from in situ
dissipation at $T \sim 10^8 \K$ (appropriate for transient magnetars in their low states) 
as opposed to $5-10\times 10^8 \K$ (appropriate for persistently bright
magnetars).   The contribution of dripped neutrons to the specific
heat varies strong with density:  it is strongly suppressed by superfluidity where the pairing temperature is high
compared with $T$.   Runaway creep is therefore concentrated within the superfluid part of the lower crust.

Two terms are added to the right-hand side of Equation (\ref{eq:en}), representing thermal conduction out of the 
`spot' over a characteristic size (temperature gradient scale) $\ell$; and neutrino cooling.   The thermal conductivity
varies weakly with temperature near $10^{14}$ g cm$^{-3}$, $K_{\rm cond} \simeq 1\times 10^{20}$ erg (cm-s-K)$^{-1}$
\citep{potekhin15}.  Neutrino emission is dominated by thermal bremsstrahlung and synchrotron radiation by relativistic
electrons, and calculated using formulae given in \citep{yakovlev01}.
The conductive cooling length is taken to be $\ell = 0.3$ km, approximately
the vertical density scale height at the base of the crust.  

Here ohmic effects can be neglected, in contrast with the microscopic fault model of \cite{LL12}, which assumes that the
fault width is much smaller than $\delta R_c$.  Both that analysis and the magneto-thermal front model of 
\cite{bl14} fail to include a component of the stress of non-local origin.  As a result, these authors find
that dissipation growth depends on sharp gradients in temperature and Lorentz force.  

A simple numerical experiment evolves the `spot' starting from an initial temperature $T_{\rm base}$, with the 
initial total stress $\sigma_0 = \sigma_{g,0} + \sigma_{l,0}$ corresponding to a very long creep time ($> 1$ Myr) 
at this temperature.  The transverse magnetic field is normalized by the parameter $f_{\rm local}$ in Equation
(\ref{eq:locsig}), and
we consider a range of $t_{\rm growth}$ for the global stress ranging from about a day up to $10^5$ yr 
(longer than a typical magnetar lifetime).  

Figure \ref{fig:zeroD} shows the evolution of creep rate, total stress, and temperature in the spot.  As the applied
stress ramps up more quickly, one finds not surprisingly a faster and sharper response of the plastic flow in the spot.  
The creep rate $\dot\varepsilon_{\rm pl}$ peaks and then subsides as the transverse magnetic field is sheared out, which allows
the local Maxwell stress to cancel out further increase in the global stress.  Smaller initial temperatures correspond
to higher peak creep rates, a consequence of the $T^3$ scaling of the specific heat.  

The stress, temperature and creep rate all eventually reach a plateau.  This represents a balance between in situ
plastic heating and neutrino cooling for all except the longest $t_{\rm growth}$ and the lower values of $T_{\rm base}$.  
One observes some bimodality in final temperature and stress in these latter cases, representing a switch between
a warmer, neutrino-cooled state and a cooler state in which plastic heating is balanced by conductive cooling.
Large swings in creep rate are observed in a few cases where $t_{\rm growth}$ is comparable to 1 yr, the conductive
cooling time across $\ell \sim 0.3$ km.  

\begin{figure}
\vskip -0.5in
\epsscale{1.2}
\plotone{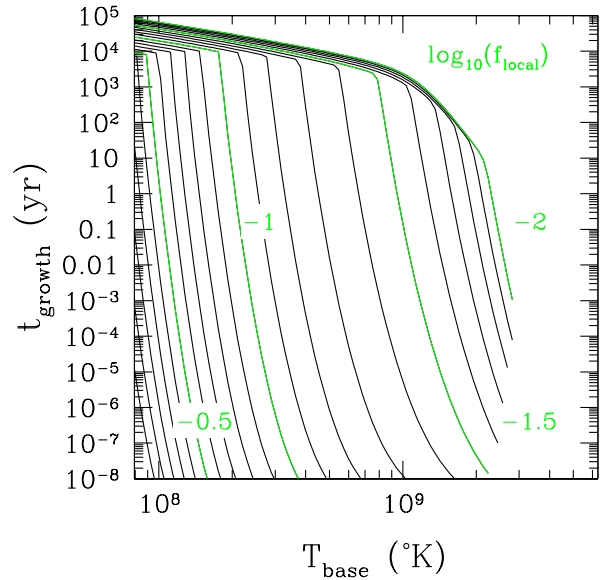}
\vskip -0.8in
\caption{Critical value of the growth time $t_{\rm growth}$ of the global stress that is applied to a small
plastic spot in a neutron star crust, above which the spot experiences runaway plastic creep reaching
a peak rate $\sim 0.1 V_\mu/\delta R_c \sim 300$ s$^{-1}$.  This is plotted versus the baseline temperature 
of the spot at the onset of heating, for a range of values of the magnetic field strength transverse to the
direction of plastic creep.  The curves are uniformly spaced in ${\rm log}_{10}f_{\rm local}$ between -2 and 0
(see Equation \ref{eq:locsig}).}
\vskip .1in
\label{fig:critgrowth}
\end{figure}

One infers from Figure \ref{fig:zeroD} that there is a critical value of $t_{\rm growth}$ for the applied stress
above which $\dot\varepsilon_{\rm pl}$ approaches the dynamical value $V_\mu/\ell$, where inertial forces must be taken
into account,
and the flow rate must saturate.  Here $V_\mu = (\mu/\rho)^{1/2}$ is the shear wave speed in an unmagnetized Coulomb solid.  
Figure \ref{fig:critgrowth} shows that the crust is most susceptible to runaway creep at lower temperatures:  
slower growth of the applied stress (higher $t_{\rm growth}$) is required.   Alternatively, strengthening the
transverse magnetic field makes runaway creep more difficult.  Indeed, at fixed $T_{\rm base}$ one can consider there
to be a threshold value of $f_{\rm local}$ leading to explosive creep growth.  For example, at $T_{\rm base} = 
5\times 10^8$ K the critical $t_{\rm growth}$ drops from $10^{11}$ s (comparable to the magnetar lifetime) down to 
$\sim 1$ s as $f_{\rm local}$ rises from $0.05$ to $0.1$.

\subsection{Plastic Response in One Dimension: \\ Conductive Fault with Imposed Global Stress}

We now turn to a steady, 1-dimensional model of plastic creep in a fault-like geometry.  The injection of heat
by plastic creep is balanced by volumetric neutrino cooling and thermal conduction away from the maximum of 
temperature and stress.  Similar solutions were constructed some time ago by \cite{schubert78}, with a goal of
explaining the formation of faults in the Earth's lithosphere.   

The vertical scaleheight $\delta R_c$ provides a characteristic scale below which the temperature gradient 
is mainly horizontal, and vertical conductive losses can be neglected, so that $T = T(x)$.   We first consider
a weakly-magnetized fault, which is possible when the solid stress is sourced by long-range imbalances in 
the Maxwell stress at the crust-core interface (Figure \ref{fig:global_stress}).

We construct steady profiles of temperature, strain rate and heat flux, with a range of background temperature 
$T(\delta R_c) = T_{\rm base}$, from the coupled equations
\ba
F_{\rm cond} &=& -K_{\rm cond}{dT_{\rm cond}\over dx}; \nn
\sigma \dot\varepsilon_{\rm pl}(T,\sigma) &=& -{dF_{\rm cond}\over dx} - \dot Q_\nu(T,B).
\ea
Each value of $T_{\rm base}$ defines a sequence of solutions with different
peak temperatures $T_{\rm mid} \equiv T(0)$ at the middle of the fault, and different total stress
$\sigma = \sigma_g =$ const.   The conductive energy flux $F_{\rm cond}$ vanishes at $x = 0$, corresponding 
to a solution that is symmetric about the temperature maximum: $T(-x) = T(x)$ and $F_{\rm cond}(-x) = -F_{\rm cond}(x)$.  

\begin{figure}
\vskip -0.5in
\epsscale{1.2}
\plotone{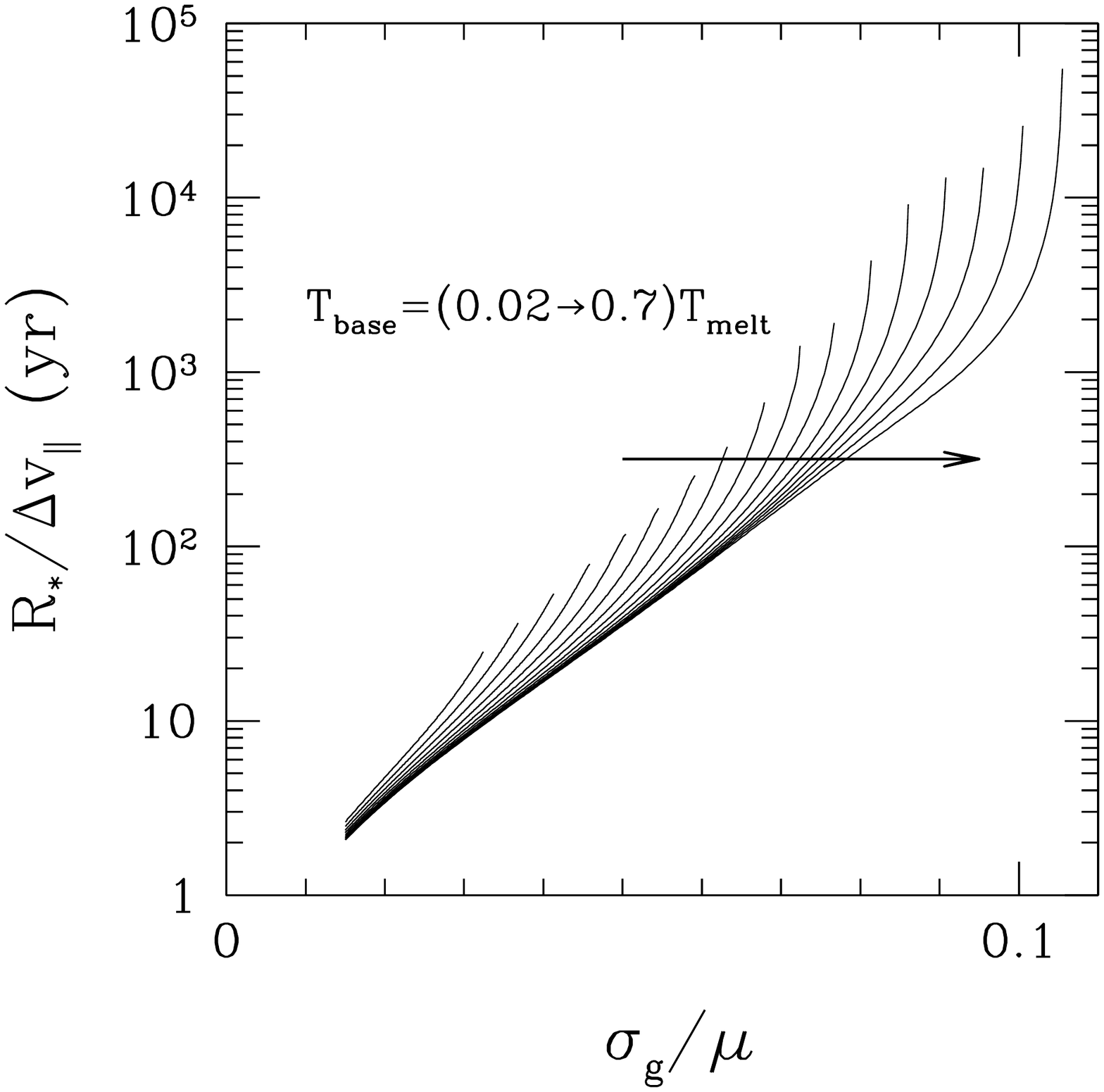}
\vskip -1.2in
\plotone{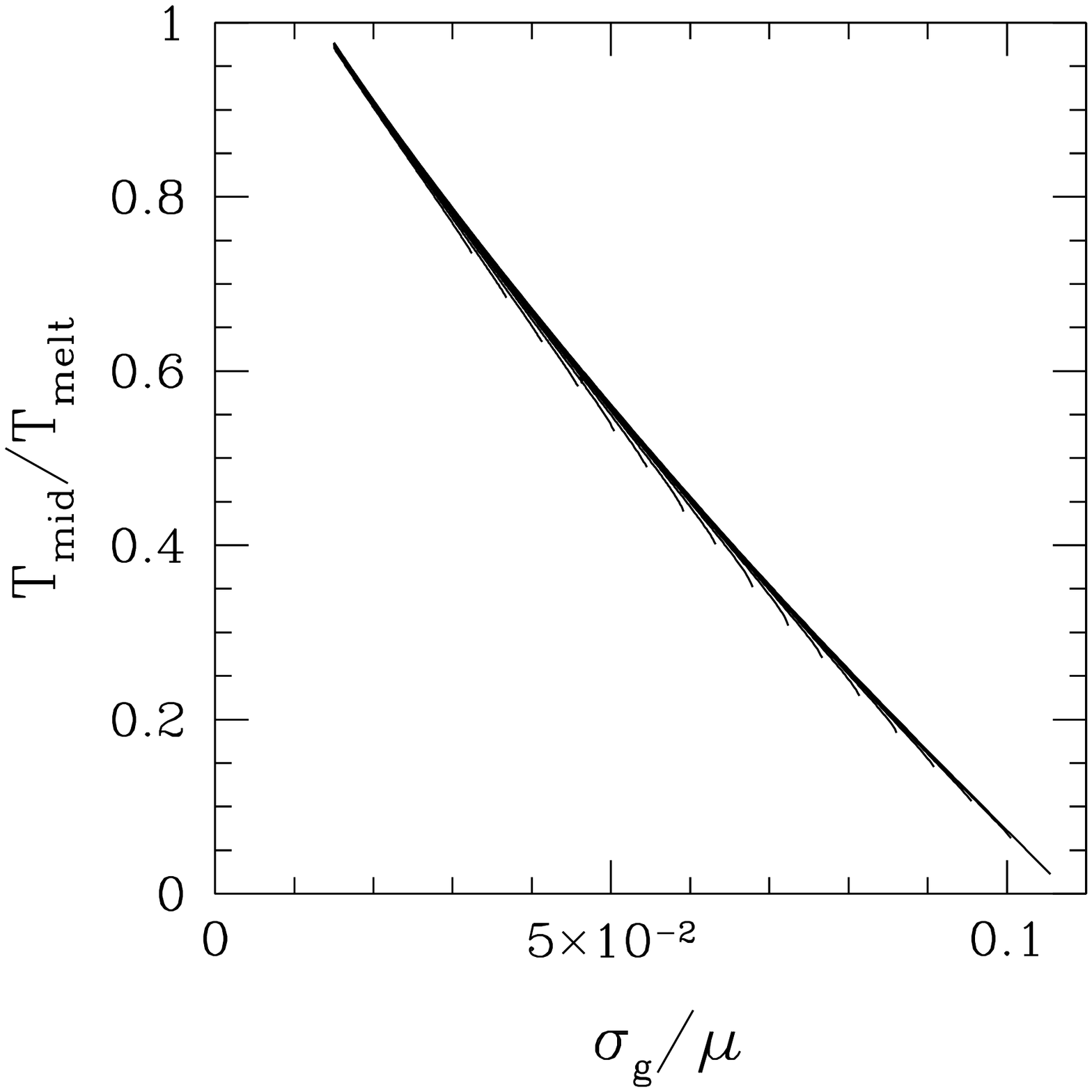}
\vskip -0.65in
\caption{{\it Top panel:} Global creep time (\ref{eq:tcreep}) in the neutron star crust resulting from plastic flow at
a narrow fault, versus the magnitude of the global stress that is applied to the fault.  Here the
magnetic field transverse to the fault direction is assumed to vanish.  Curves correspond to a range
of background temperatures outside the fault.  {\it Bottom panel:} peak temperature in the fault, again for a 
range of background temperatures.}
\vskip .2in
\label{fig:oneD}
\end{figure}

The global stress relaxes on a timescale 
\be\label{eq:tcreep}
t_{\rm creep} \sim {R_\star\over \Delta v_\parallel},
\ee
where 
\be\Delta v_\parallel = \int_0^{\delta R_c} \dot\varepsilon_{\rm pl}[T(x),\sigma] dx
\ee
is the net differential creep speed across the fault.  
Figure \ref{fig:oneD} shows the sequence of curves relating $t_{\rm creep}$ to the magnitude of the stress.   
Each curve is labeled by $T_{\rm base}$, with lower stresses corresponding to higher mid-plane temperatures.   
One also sees a tight relation between $T_{\rm mid}$ and $\sigma$, reflecting the 
extreme sensitivity of $\dot\varepsilon_{\rm pl}$ to $\sigma$ and $T$.  

\section{Elastic-Plastic-Thermal Evolution of Neutron Star Crust:  Two Dimensional Model}\label{s:three}

We now construct a global model of yielding in neutron star crusts.  Stresses are redistributed horizontally
in response to an imbalance in the Maxwell stress at the crust-core boundary relative to the magnetar surface.
An effectively 2-dimensional description is made possible by the small scale height $\delta R_c \sim 0.03\,R_\star$
at the base of the crust.  
Our focus is on the feedback of the global 2-dimensional stress field on localized yielding. 

We first review some basic energetic considerations related to large-scale crustal yielding, and describe
some technical details.  Section \ref{s:four} presents results of time-dependent calculations, and
Section \ref{s:five} the implications for magnetar giant flares.

\subsection{Slow Evolution of the Core Magnetic Field away from Magnetohydrostatic Equilibrium}

The magnetic field within the magnetar core is slowly reconfigured by ambipolar drift and beta reactions
\citep{Haensel90,GR92,Pethick92} over its active lifetime of $t_{\rm NS} \sim 10^{11}$ s.  The feedback
of heating on the rate of modified URCA reactions allows the diffusion time to scale with the lifetime of the star
while the core neutrons remain in a normal state \citep{TD1996}.  Slow transport of the magnetic field through a
fluid star allows the field to evolve into a non-axisymmetric configuration \citep{braithwaite08} that becomes susceptible 
to a global hydromagnetic instability.   

Many details of how this happens cannot yet be captured by global hydromagnetic simulations,
in which numerical diffusion is typically rapid in the outer parts of the model star.  Nonetheless, simple considerations
of continuity suggest that modes of a very low growth rate must develop during a gradual transition away from
stability.  We make the simplest approximation of treating the core field as a 1-dimensional spring, 
with a spring constant $\propto \omega_0^2$.  Stable equilibrium corresponds to  $\omega_0 \sim V_{\rm A}/R$,
where $V_{\rm A}$ is the Alfv\'en speed within the core.

Now we let the spring constant evolve linearly through zero over the diffusion time $\sim t_{\rm inst}$
(which initially may be comparable to $t_{\rm NS}$),
\be
   \omega_0^2 \sim  \left({V_{\rm A}\over R}\right)^2\left(1 - {t\over t_{\rm inst}}\right).
\ee
Near $t = t_{\rm inst}$, one finds $|\omega_0|$ is much smaller than the spin frequency $\Omega$ (typically $\sim 1$ s$^{-1}$
for an X-ray bright magnetar), meaning that the growth of a hydromagnetic instability is limited by the Coriolis force.
Balancing this force against internal magnetic stresses gives $\Omega |v| \sim |\omega_0^2| R_\star$.  
The hydromagnetic displacement at speed $v$ can then be converted to a growth rate
\be
\Gamma_{\rm eff} \sim -{\omega_0^2\over\Omega}.
\ee
Instability occurs where $\omega_0^2 < 0$, but in a fluid star it can develop only when $\Gamma_{\rm eff} \Delta t > 1$,
where $\Delta t = t-t_{\rm inst}$.   This gives a minimum growth time
\be
\Gamma_{\rm eff}^{-1} \gtrsim (R/V_{\rm A})^{1/2} t_{\rm inst}^{1/2},
\ee
which is much shorter than the neutron star lifetime, but still longer than the Alfv\'en crossing time 
$R/V_{\rm A} \sim 0.1$ s.  Taking $t_{\rm inst} \sim t_{\rm NS} \sim 10^{11}$ s gives
$\Gamma_{\rm eff}^{-1} \sim 10^5$ s.

These considerations apply to a fluid star.  The next question which arises 
is whether the crustal elastic stress superposed on the fluid core should force the 
transition from a decaying to a growing mode away from vanishing core spring
constant ($\omega_0 = 0$).   That would be the case if, at each instant in the
evolution of the star, the core could be placed in magneto{\it hydrostatic} equilibrium, with
vanishing shear component of the Maxwell stress at the base of the crust.
But that is clearly not the case if we start the magnetar at some early time $t$ with the 
shear stress above the yield stress.   When the core is otherwise close to magnetohydrostatic equilibrium,
the shear Maxwell stress can only relax to the point where the plastic creep rate has dropped to
$\dot \varepsilon \sim \varepsilon_b/t$.  Here $\varepsilon_b \equiv \sigma(\dot\varepsilon_{\rm pl})/\mu \sim 
0.02$-0.1 is the yield strain at creep rate $\dot\varepsilon_{\rm pl}$ as described by Equations (\ref{eqy1})
and (\ref{eqy}).

\subsection{Response of the Crust to a Hydromagnetically Imbalanced Core Magnetic Field}

This suggests a numerical experiment in which the creep rate starts at a very low value
in the plastic parts of the crust.  In the initial state, the lattice strain exceeds
the yield strain only in small patches.  We allow the Maxwell stress to grow very slowly,
and then see if the crust is capable of a much faster response.  Only then we vary the sign of the core 
spring constant.

We idealize the crust as a planar solid layer that is threaded by a uniform vertical magnetic field $B_z$,
which is tilted by some angle at the lower boundary (Figure \ref{fig:global_stress}).  The horizontal
magnetic field at the lower boundary can be expressed in terms of a solenoidal displacement field $\bxi_B$ at
the base of the core (depth $h_{\rm core} \sim R_\star$)
\be 
{\bf B}_\perp^- = -{\bxi_B\over h_{\rm core}} B_z,  \quad i = x,y,
\ee
with $\partial_x B_x^- + \partial_y B_y^- = \partial_x\xi_{B,x} + \partial_y\xi_{B,y} = 0$.  We define
$\bxi_B$ as the net displacement before any compensating horizontal motion of the crust.  The applied Maxwell stress drives both elastic and plastic 
displacements in the crust,
\be\label{eq:xi}
\bxi(x,y) = \bxi_{\rm el}(x,y) + \bxi_{\rm pl}(x,y),
\ee
in response to which
\be\label{eq:db}
{\bf B}_\perp^- \rightarrow {{\bxi - \bxi_B}\over h_{\rm core}}B_z.
\ee

Since the crustal shear modulus is only $\sim 10^{-3}$ of the hydrostatic pressure, the displacement field
(\ref{eq:xi}) is  nearly 2-dimensional and incompressible.  Therefore 
both $\bxi_B$ and $\bxi$ can be expressed using stream functions $\Psi_B$, $\Psi
 = \Psi_{\rm el} + \Psi_{\rm pl}$,
with (e.g.) $\xi_i = \epsilon_{ij} \partial_j\Psi$.   Here $\epsilon_{ij} = \pm 1$ is the antisymmetric symbol.

In this approximation, the wavelength of horizontal variations 
in $\Psi_B$ and $\Psi$ is constrained to be larger than or 
comparable to $\delta R_c$.   
We neglect any vertical structure in $\bxi$, that is, magnetic tilt interior to the crust,
whose evolution will be considered in Section \ref{s:six}.   

In some calculations, we also include a magnetic field localized to the crust.
The seed horizontal magnetic field ${\bf B}_{\perp,0}^c$ is uniform and configured 
independently of the core field.   Elastic and plastic displacements of the crust
perturb this field and also change the tilt of the core field,
\ba
B_x^- &=& {\partial_y(\Psi - \Psi_B)\over h_{\rm core}};  \quad
B_y^- = -{\partial_y(\Psi - \Psi_B)\over h_{\rm core}};  \nn
B_x^c-B_{x,0}^c &=& B_{x,0}^c \partial_x\xi_x + B_{y,0}^c \partial_y\xi_x\nn
                &=& B_{x,0}^c \partial_x\partial_y\Psi + B_{y,0}^c \partial_y^2\Psi; \nn
B_y^c-B_{y,0}^c &=& B_{x,0}^c \partial_x\xi_y + B_{y,0}^c \partial_y\xi_y\nn
                &=& -B_{x,0}^c \partial_x^2\Psi - B_{y,0}^c \partial_x\partial_y\Psi.
\ea
A tangled component of the magnetic field is easy to implement in Fourier space, as discussed below.

The equation of magnetoelastic equilibrium reads
\be\label{eq:me}
\partial_z\left({B_i B_z\over 4\pi}\right) + \mu\partial_j\left(\sigma_{ij} + \sigma^M_{ij}\right),
\ee
where the solid stress is 
\ba\label{eq:sigsol}
\sigma_{ij} &=& \mu\left(\partial_i\xi_{{\rm el},j} + \partial_j\xi_{{\rm el},i}\right) + P\delta_{ij};\nn
\partial_j\sigma_{ij} &=& \mu\epsilon_{ik}\partial_k(\partial^2\Psi_{\rm el}) + \partial_i P,
\ea
the horizontal Maxwell stress is 
\be
\sigma^M_{xx} = {{(B^c_x)^2-(B^c_y)^2}\over 8\pi} = -\sigma^M_{yy};\quad
\sigma^M_{xy} = {B^c_x B^c_y\over 4\pi},
\ee
and
\ba
\partial_i \sigma^M_{xi} &\;=\;& {B^c_y\over 4\pi}(\partial_yB^c_x - \partial_xB^c_y)\nn
   &\;\simeq\;&  {B^c_{y,0}\over 4\pi}\left(B^c_{x,0}\partial_x\partial^2\Psi + B^c_{y,0}\partial_y\partial^2\Psi\right);\nn
\partial_i \sigma^M_{yi} &\;=\;& -{B^c_x\over 4\pi}(\partial_yB^c_x - \partial_xB^c_y)\nn
   &\;\simeq\;&  -{B^c_{x,0}\over 4\pi}\left(B^c_{x,0}\partial_x\partial^2\Psi + B^c_{y,0}\partial_y\partial^2\Psi\right).
\ea
Only the elastic displacement field enters into the solid stress (\ref{eq:sigsol}).

We integrate Equation (\ref{eq:me}) in vertical coordinate $z$ from the bottom to the top of the crust, neglecting the 
exterior\footnote{The magnetosphere generally supports only a weaker magnetic shear than the magnetar interior.}
Maxwell stress $B_i^+B_z/4\pi$, and defining vertically averaged shear modulus and pressure via
$\int \mu dz = \bar\mu  \delta R_c$, $\int P dz = \bar P \delta R_c$.  This gives
\ba\label{eq:me2}
&& {B_z^2\over 4\pi}{\partial_y(\Psi_B - \Psi_{\rm el})\over h_{\rm core}} + 
\delta R_c\biggl[\bar\mu \partial_y\partial^2\Psi_{\rm el} + \nn
&& {B^c_{y,0}\over 4\pi}\left(B^c_{x,0}\partial_x\partial^2\Psi + B^c_{y,0}\partial_x\partial^2\Psi\right)
+ \partial_x \bar P\biggr] \;=\; 0;\nn
&& {B_z^2\over 4\pi}{\partial_x(\Psi_B - \Psi_{\rm el})\over h_{\rm core}} +
 \delta R_c\biggl[\bar\mu \partial_x\partial^2\Psi_{\rm el} + \nn
 && {B^c_{y,0}\over 4\pi} \left(B^c_{x,0}\partial_x\partial^2\Psi + B^c_{y,0}\partial_x\partial^2\Psi\right)
- \partial_y \bar P\biggr] \;=\; 0.\nn
\ea
One observes that the pressure perturbation must be included only when ${\bf B}^c_{\perp,0}$ is non-vanishing.
We neglect forces arising from background gradients in ${\bf B}^c_{\perp,0}$.  

One solves for $\bar P$ by taking $x-$ and $y-$ derivatives of the first and second lines in Equation (\ref{eq:me2}),
and subtracting:
\be
\bar P = {B^c_{x,0}B^c_{y,0}\over 4\pi}(\partial_y^2-\partial_x^2)\Psi + 
      {(B^c_{x,0})^2 - (B^c_{y,0})^2 \over 4\pi} \partial_x\partial_y\Psi.
\ee
Substituting this back into Equation (\ref{eq:me2}) gives
\be
\epsilon_{ij}\partial_j \Phi = 0,
\ee
where 
\ba
&&\Phi = {B_z^2\over 4\pi}{\Psi_B-\Psi\over h_{\rm core}} + \nn
        \delta R_c\biggl[\mu\partial^2\Psi_{\rm el} +
&& {1\over 4\pi}\left(B^c_{x,0}\partial_x + B^c_{0,y}\partial_y\right)^2(\Psi_{\rm el} + \Psi_{\rm pl}) \biggr] = 0.\nn
\ea
Uniqueness of the solution implies that $\Phi$ vanishes.  

The elastic displacement stream function $\Psi_{\rm el}$ is easily solved for in Fourier space $(k_x,k_y)$ in
terms of $\Psi_B$ and the cumulative plastic stream function $\Psi_{\rm pl}$:
\ba\label{eq:psiel}
&& \Psi_{\rm el,\bf k} \;=\;
-\left[1 + \delta R_c h_{\rm core} \left({4\pi\mu\over B_z^2}k^2 + 
{({\bf k}\cdot{\bf B}^c_0)^2\over B_z^2}\right)\right]^{-1} \times \nn
&& \left\{-\Psi_{B,\bf k} + \Psi_{\rm pl,\bf k}\left[1 + \delta R_c h_{\rm core} 
{({\bf k}\cdot{\bf B}^c_0)^2\over B_z^2}\right]\right\}.
\ea
To introduce a tangled magnetic field $B_t$ with no preferred direction, we replace
$({\bf k}\cdot{\bf B}_0^c)^2 \rightarrow ({\bf k}\cdot{\bf B}_0^c)^2 + k^2B_t^2$ in 
Equation (\ref{eq:psiel}).

\subsubsection{Numerical Procedure}

Our procedure therefore is as follows.  

1. A Maxwell stress is imposed at the lower crust boundary with a power-law spectrum 
\be\label{eq:spec}
\left({B_{x,y} B_z\over 4\pi}\right)_k \propto (k_\parallel^2 + \eta k_\perp^2)^{-(1+\alpha)}
\ee
and random phases.\footnote{Strictly the vector potential $A_z$ is first calculated and
then $B_{x,y}$ are derived from it.}   This corresponds to a stress $B_xB_z/4\pi \propto k^{-2\alpha}$ 
per logarithm of wavenumber.  The parameter $\eta$ can be chosen to be less than unity to represent an
anisotropic Maxwell stress, e.g. a nearly axially symmetric core field with predominantly latitudinal
gradients.   The basis $(k_\parallel,k_\perp)$ can also be rotated with respect to $(k_x,k_y)$:  we
typically choose a $45^\circ$ rotation when $\eta < 1$, so as to eliminate the possibility of grid-alignment
effects.  

2. The magnitude of the applied Maxwell stress is gradually raised until yielding begins at
an appreciable rate.  We adaptively test the time step, raising or reducing it so as to 
maintain a maximum fractional change $\sim 10^{-2}$ in the creep rate in any pixel.  As a result,
the timestep can vary from $\sim 10^3$ yr or longer down to a fraction of a millisecond.  The
creep rate, as determined by Equation (\ref{eqy1}), is capped at $0.1 V_\mu / \delta R_c$, representing
the feedback of inertial forces on unbalanced stresses.

3. In each timestep $\delta t$, the plastic stream function is updated according to the method
described in Section \ref{s:update}:  $\Psi_{\rm pl, \bf k} \rightarrow \Psi_{\rm pl,\bf k} + 
\dot\Psi_{\rm pl,\bf k}\delta t$, where $\dot\Psi_{\rm pl,\bf k}$ is given by Equation (\ref{eq:dotPsi}).
We also allow the applied Maxwell stress to adjust to plastic creep:
the core magnetic stream function is updated according to 
$\Psi_B \rightarrow \Psi_B \mp \dot\Psi_{\rm pl, \bf k}\delta t$.  Then the elastic response of the 
crust is recalculated according to Equation (\ref{eq:psiel}).  The temperature is updated according to
Equation (\ref{eq:en}).

4. The effective spring constant in the core can be taken to be positive or negative, corresponding
to the upper or lower sign in the update for $\Psi_B$.  The change in core field due to creep
supplements the imposed linear rise in the Maxwell stress, and mediates a runaway global instability
when the core spring constant is negative.  

5. We separately test the presence or absence of a horizontal crustal magnetic field.  In particular,
a tangled crustal field has a strong effect on suppressing runaway yielding (Section \ref{s:four}).

\subsection{Formulation of Plastic Flow in a Compact Zone}\label{s:update}

Here we make use of the {\it ab initio} relation $\sigma_b(\dot\varepsilon_{\rm pl})$ (Equation (\ref{eqy1})) between 
breaking strain and strain rate to describe inhomogeneous plastic creep in one and two dimensions.
The generalization to fully three-dimensional deformations is not addressed.

The molecular dynamics simulations of 3-dimensional solids on which this is based
take the imposed shear to have planar symmetry and grow at a uniform rate, e.g. $\varepsilon = 
\dot\varepsilon_{\rm pl} t$ with $\dot\varepsilon_{\rm pl} = \partial_x v_y = $ const \citep{Chugunov2010,hoffman12}.

The plastic strain rate $\dot\varepsilon_{\rm pl}$ is defined as a positive scalar function 
of the scalar stress $\sigma$.  In the planar model described here, it is given by
\be
\sigma = \mu\varepsilon \equiv 2\mu(\varepsilon_{xx}^2 + \varepsilon_{xy}^2)^{1/2}
\ee
where 
\be\label{eq:2Dstrain}
\varepsilon_{xy} = {1\over 2}(\partial_y^2- \partial_x^2)\Psi_{\rm el};  \quad 
    \varepsilon_{xx} = -\varepsilon_{yy} = \partial_x\partial_y\Psi_{\rm el}.
\ee

The evolution of the strain tensor by plastic creep is described by a single scalar
function $\dot \Psi_{\rm pl}$.  Here we treat all parts of the crust as effectively plastic, but
with exponentially varying creep rate.  As long as the deformation rate is low enough
that inertial forces can be neglected, the crust moves through a series of quasi-equilibrium
states.  Then the new elastic equilibrium can be obtained from Equation (\ref{eq:psiel})
after the plastic stream function is updated to  $\Psi_{pl} \rightarrow \Psi_{\rm pl} + \dot\Psi_{\rm pl} \delta t$.

\subsubsection{Inhomogeneous One-dimensional Creep and \\ Crack Formation}

Some subtleties are involved in obtaining a self-consistent time-evolution equation for $\Psi_{\rm pl}$
in a compact domain.  Consider first the case of linear creep in cylindrical symmetry with one periodic coordinate,
\be
\bxi({\bf x}) = \xi_y(x) \hat y;  \quad  \xi_y(x+L) = \xi_y(x).
\ee
At first it might appear straightforward to apply the continuous creep law (\ref{eqy1}) to the 
stress profile $\sigma = \mu |\xi'_{y,\rm el}|$:
\be\label{eq:cylcreep}
\dot\xi_y' = {\rm sgn}(\xi_{y,\rm el}')\dot \varepsilon(\varepsilon_{\rm el})
\ee
where $\dot\varepsilon = \dot\varepsilon_{\rm pl} + \dot\varepsilon_{\rm el}$
and $\varepsilon_{\rm el} \equiv |\xi_{y,\rm el}'|$.  
But an obstruction arises from the fact that the net displacement
\be\label{eq:delxi}
\Delta\xi_y = \int_0^L \xi_y' dx = {\rm const}
\ee
must be invariant under continuous changes in $\xi_y$, whereas the integral
\be
\dot\Delta\xi_y = \int_0^L {\rm sgn}(\xi_{y,\rm el}')\,\dot\varepsilon dx \neq 0.
\ee
does not vanish in general. 

Equivalently, Equation (\ref{eq:cylcreep}) does not prescribe the
time evolution of the zero-frequency mode of $\Psi$.  Fourier transforming,
\be
\Psi = \sum_{r = -N/2}^{N/2-1} \Psi_r e^{i2\pi r x/L},
\ee
Equation (\ref{eq:cylcreep}) is equivalent to
\be\label{eq:dpfour}
-r^2\dot\Psi_{{\rm pl},r} = \left[{\rm sgn}(\partial_x^2\Psi_{\rm el})\dot \varepsilon_{\rm pl}\right]_r
     = -{1\over N}\sum_{r = -N/2}^{N/2-1} l^2 \Psi_{{\rm el},l} 
\left({\dot\varepsilon_{\rm pl}\over\varepsilon_{\rm el}}\right)_{r-l}.
\ee

This conundrum exists only if the deformation is continuous.   Introducing a dislocation near the
location of peak strain would remove (\ref{eq:delxi}) as a conserved quantity and restore consistency with
the local creep law (\ref{eq:cylcreep}).  

\subsubsection{Self-consistent Two-dimensional Creep}

A different constraint on the creep law is present in a 2-dimensional solid with inhomogeneous deformation.
Here it is natural to consider aligning the plastic deformation tensor with the stress tensor \citep{hill98}
\be\label{eq:2Dcreep}
\dot\varepsilon_{ij,\rm pl} = \varepsilon_{ij,\rm el}{\dot\varepsilon_{\rm pl}\over\varepsilon_{\rm el}}.
\ee
However, the strain tensor is derived from an incompressible displacement field and must, therefore,
satisfy consistency rules.  Equation (\ref{eq:2Dstrain}) implies that
\be\label{eq:consis}
{1\over 2}\left(\partial_x^2 -\partial_y^2\right) \varepsilon_{xx,\rm pl} 
       = \partial_x\partial_y \varepsilon_{xy,\rm pl}.
\ee
The update $\delta\varepsilon_{ij,\rm pl}  = \dot\varepsilon_{ij,\rm pl}\delta t$ obtained from 
Equation (\ref{eq:2Dcreep}) over a small time interval $\delta t$ 
will satisfy Equation (\ref{eq:consis}) when $\dot\varepsilon_{\rm pl}$ is constant,
and also when the strain field has a planar or rotational symmetry, but not more generally.  

A self-consistent approach is obtained by working with the stream function $\Psi_{\rm pl}$ and in Fourier
space,
\be
\Psi_{\rm pl}(x,y) = \sum_{r,s = -N/2}^{N/2-1} \Psi_{{\rm pl}\,r,s} e^{2\pi i r (x/L)} e^{2\pi i s (x/L)}.
\ee
The following relation reduces to Equation (\ref{eq:dpfour}) when the strain field
has a planar symmetry (e.g. when only $\varepsilon_{xy}$ or $\varepsilon_{xx} = -\varepsilon_{yy}$,
or some constant linear combination of these components, is non-vanishing throughout the solid):
\ba\label{eq:dotPsi}
\dot\Psi_{{\rm pl}\,r,s} &=& {2rs\over (r^2+s^2)^2}
\sum_{l,m} 2lm\Psi_{lm,\rm el}\left({\dot\varepsilon_{\rm pl}\over\varepsilon_{\rm el}}\right)_{r-l,s-m} +\nn
&&{r^2-s^2\over (r^2+s^2)^2}
\sum_{l,m} (l^2-m^2) \Psi_{lm,\rm el}\left({\dot\varepsilon_{\rm pl}\over\varepsilon_{\rm el}}\right)_{r-l,s-m}.
\ea
Similarly to the 1-dimensional example discussed above, this equation does not evolve the $(r=0,s=0)$ 
Fourier mode.

The 2-dimensional description of the strain pattern must break down on lengthscales shorter than the crust
scale height $\delta R_c$, and so we impose a cutoff $N \sim R_\star/\delta R_c$ in Fourier space.

\begin{figure}
\vskip -0.5in
\epsscale{1.15}
\plotone{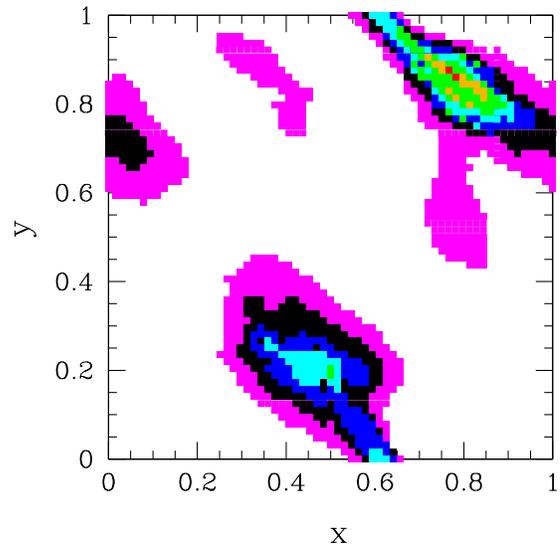}
\vskip -0.7in
\caption{Two-dimensional pattern of yielding in a planar Coulomb solid of characteristic density
$10^{14}$ g cm$^{-2}$ that is threaded with a vertical magnetic field $B_z = 10^{15}$ G and 
subjected to Maxwell stress $B_\perp B_z/4\pi$ from below.  Imposed stress is
mildly anisotropic  (parameter $\eta = 0.3$ in
Equation (\ref{eq:spec})) with a $45^\circ$ tilt from the grid.  Grid $2^6$ cells on a side.  
In this realization, the core magnetic spring constant is taken to be negative, so that
the applied Maxwell stress grows exponentially in time in response to plastic motion in the crust.
Mauve, black, blue, cyan, green, gold, and red points show the 
growth of the zones with creep rate exceeding 
$10^{-5}$ s$^{-1}$.  Figure \ref{fig:2Danis} shows for comparison the result of long-term
evolution with a hydromagnetically stable (but imbalanced) core magnetic field.  In that case
much more prominent fault-like structure emerges during the long-term plastic-elastic-thermal response of the crust.}
\vskip .2in
\label{fig:2Danis2}
\end{figure}

\begin{figure}
\epsscale{1.3}
\vskip -0.7in
\plotone{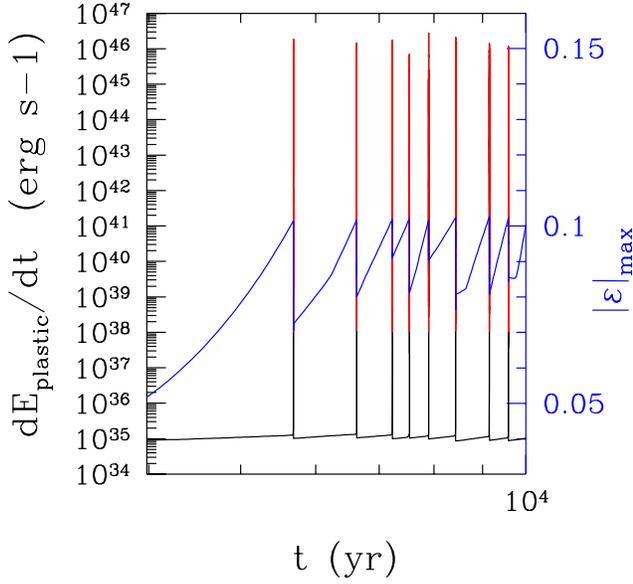}
\vskip -0.7in
\caption{`Light curve' of the integrated plastic dissipation rate in a 2D model crust, for
the same power spectrum of applied Maxwell stress that gives rise to the yielding pattern shown
in Figures \ref{fig:2Danis} and \ref{fig:2Danis2}.  Core-crust Maxwell stress has growth time 
$2000$ yr.  Here the embedded crustal magnetic field
is set to zero, so the global solid stress is not compensated by local stretching of the
magnetic field.  As a result, the first pixel to reach the yielding threshold experiences
runaway creep, with the stress change being redistributed throughout the crust and leading to 
rapid yielding at distant sites.  The yielding pattern shown in Figure \ref{fig:2Danis} corresponds
to the sixth peak in this light curve.}
\vskip .2in
\label{fig:plast_diss}
\end{figure}

\begin{figure}
\vskip -0.3in
\epsscale{1.2}
\plotone{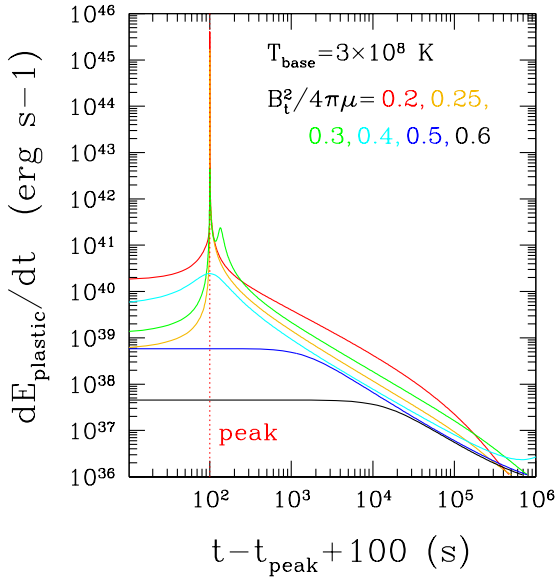}
\vskip -0.7in
\caption{Outburst behavior in 2D yielding model with embedded tangled magnetic field of varying
strengths, and $T_{\rm base} = 3\times10^8\,$K.}
\vskip .2in
\label{fig:burst_diss_Bt}
\end{figure}

\begin{figure}
\vskip -0.5in
\plotone{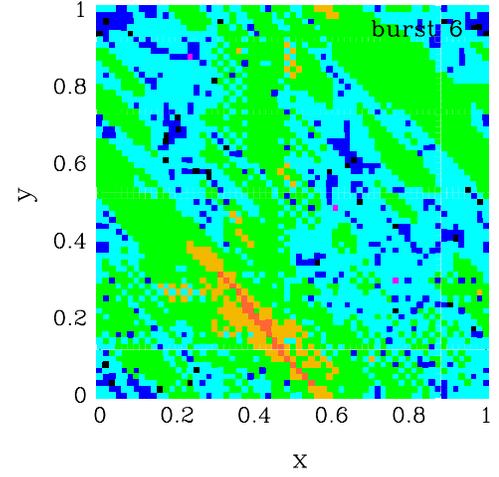}
\vskip -1in
\plotone{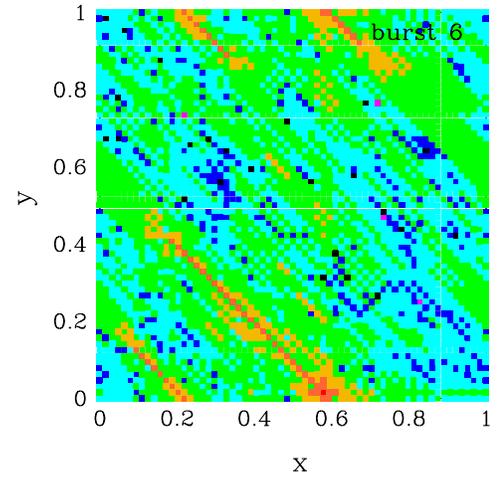}
\vskip -1in
\plotone{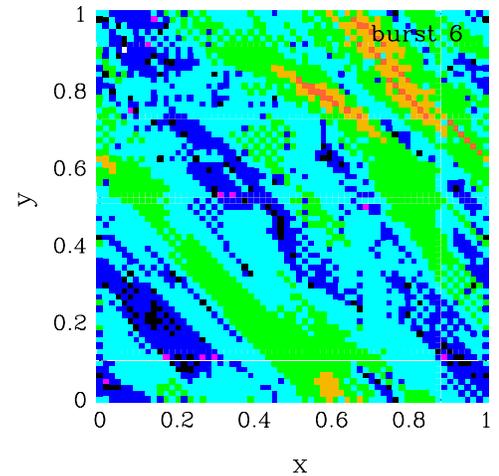}
\vskip -0.7in
\caption{Comparison of yielding pattern with different strengths and orientation of a uniform
(e.g. toroidal) crustal magnetic field $B_c$.  Here $T_{base} = 2\times 10^8 \K$ and applied Maxwell
stress is moderately anisotropic:  $\eta = 0.1$ in Equation (\ref{eq:spec}), corresponding to
a kinked core toroidal magnetic field.  {\it Top panel:}  $B_c = 0$.   {\it Middle panel}:
$B_c^2/4\pi = 0.1\,\mu$ and both crustal field and dominant symmetry direction of core field
oriented $45^\circ$.  {\it Bottom panel}:  Same $B_c$ but now pointing in the $x$-direction,
at $45^\circ$ with respect to the core field.}
\vskip .2in
\label{fig:Btor}
\end{figure}

\section{Results for the Two Dimensional Model}\label{s:four}

We now examine how the timescale of yielding, and the energy released, depend on the applied core
Maxwell stress, as well as on compensating stresses arising from the shearing of a crustal magnetic field. 
What circumstances lead to outbursts similar to giant magnetar flares?  

A typical applied Maxwell stress pattern as adopted in our calculations is depicted in Figure \ref{fig:2Danis2}.
Here the core `spring constant' is taken to be negative, so that the stress rapidly runs away
once yielding begins.   A sequence of snapshots reveals the growing area of the zones where the creep
rate exceeds a critical value $10^{-5}$ s$^{-1}$.   

Much more elaborate evolution is possible when the core spring constant is 
positive and the Maxwell stress is slowly raised at the core-crust interface.  The results of the
lower-dimensional yielding models presented in Section \ref{s:two} suggest that runaway creep
will result in the absence of a tangled crustal magnetic field. 

Figure \ref{fig:plast_diss} shows that the crust passes through a series of rapid creep events,
with energies approaching those of magnetar giant flares;  their properties are discussed in more detail
in Section \ref{s:five}.  Here the applied Maxwell stress grows on a 2000 yr timescale, and has the same
power spectrum as assumed in Figures \ref{fig:2Danis} and \ref{fig:2Danis2}.   The dissipation pattern
shown in Figure \ref{fig:2Danis} corresponds to the sixth outburst in Figure \ref{fig:plast_diss}.  The
separation between outbursts is reminiscent of the rough estimate of $40$ yr$^{-1}$ that is obtained from 
the detection of 3 giant flares from the 3 most active Galactic magnetars over the last $\sim 40$ yr. 

Including an embedded magnetic
field allows the magnetar crust to sustain larger zones with moderately fast creep, over intervals much
longer than a giant flare.   The strength of the non-linear response was found to depend on background
temperature $T_{\rm base}$, before the onset of plastic heating.   Figure \ref{fig:burst_diss_Bt} shows the 
result of introducing a tangled (statistically isotropic) magnetic field to the crust in the 2D model.
Here $T_{\rm base} = 3\times 10^8$ K, and we choose embedded field strengths ranging from 
$B_t^2/4\pi = 0.2\,\mu$ up to  $0.6\,\mu$.  A giant flare-like outburst results when $B_t^2/4\pi
\sim (0.2-0.25)\mu$, whereas slower outbursts more similar to those of AXPs are observed for the strongest
chosen values of the embedded field.

These results, along with those of Section \ref{s:two}, suggest that a wide range of outburst behavior
can result from different magnetars with only slight different strengths and configurations of magnetic
field, and slightly different temperatures.   For example, lowering $T_{\rm base}$ to $2\times 10^8$ K
in the preceding calculation broadens the transition from rapid to slow outburst behavior, as measured
by the range of $B_t$ over which this transition occurs.  Increasing $T_{\rm base}$ to $6\times 10^8$ K
makes the transition very sharp.  

The effect of an ordered (e.g. toroidal) magnetic field $B_c$ on the yielding pattern is shown in Figure
\ref{fig:Btor}.  Here we compare the result for i) vanishing $B_c$; ii)  $B_c^2/4\pi = 0.1\mu$ and
direction of $B_c$ oriented with the symmetry direction of the applied Maxwell stress (e.g. the direction
$\parallel$ in Equation (\ref{eq:spec});  and iii) same strength of $B_c$ but now oriented $45^\circ$
with respect to the $\parallel$ direction.  Fault-like features form in all three cases, with increasing
linearity when the crustal field is present and aligned with the core field.  

\begin{figure}
\vskip -0.4in
\epsscale{1.15}
\plotone{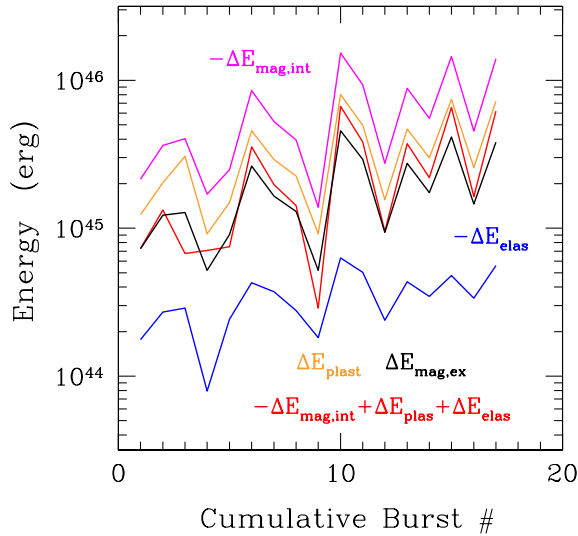}
\vskip -0.7in
\caption{Relative size of different components of the energy during a sequence of outbursts drawn 
from the same 2-dimensional simulation as the burst in Figure \ref{fig:2Danis} (burst 6).  
Curves show i) change $\Delta E_{\rm mag,int} < 0$ in internal magnetic energy (magenta);  ii) energy 
$\Delta E_{\rm plast}$ dissipated plastically in the crust (gold);  iii) energy $\Delta E_{\rm mag,ex}$
injected into the external magnetic field (black);  and iv) change in crustal elastic energy $\Delta E_{\rm elas}$
(blue) during a sequence of outbursts drawn from the same 2-dimensional simulation as the burst
in Figure \ref{fig:2Danis} (burst 6).  Parameters $B_z = 10^{15}$ G and $\eta = 0.3$ in Equation 
(\ref{eq:spec}),  with growth time 2000 yr for the applied Maxwell stress at the lower crust boundary.}
\vskip .2in
\label{fig:Eburst}
\end{figure}

\begin{figure}
\vskip -0.3in
\epsscale{1.05}
\plotone{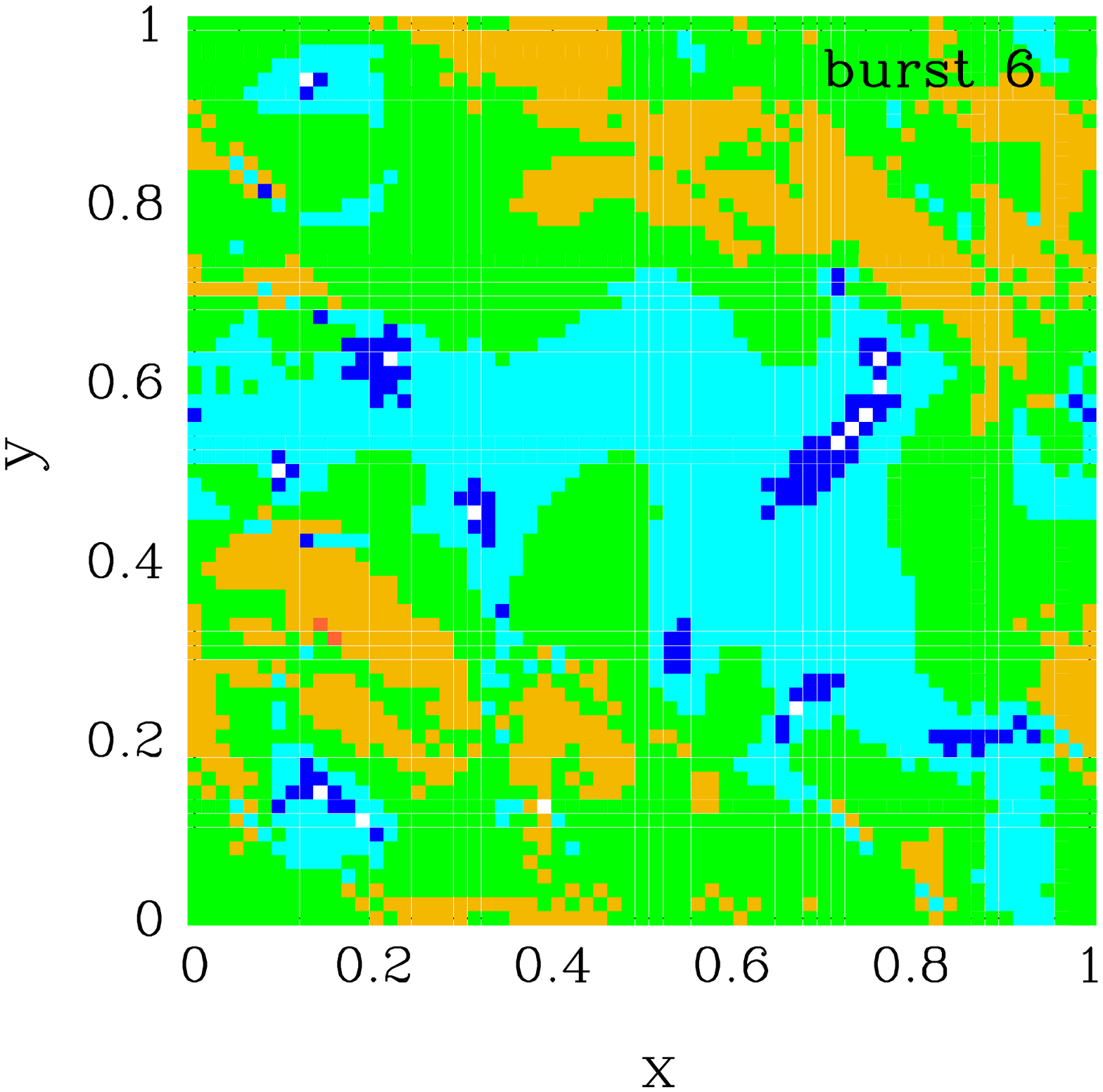}
\vskip -1.1in
\plotone{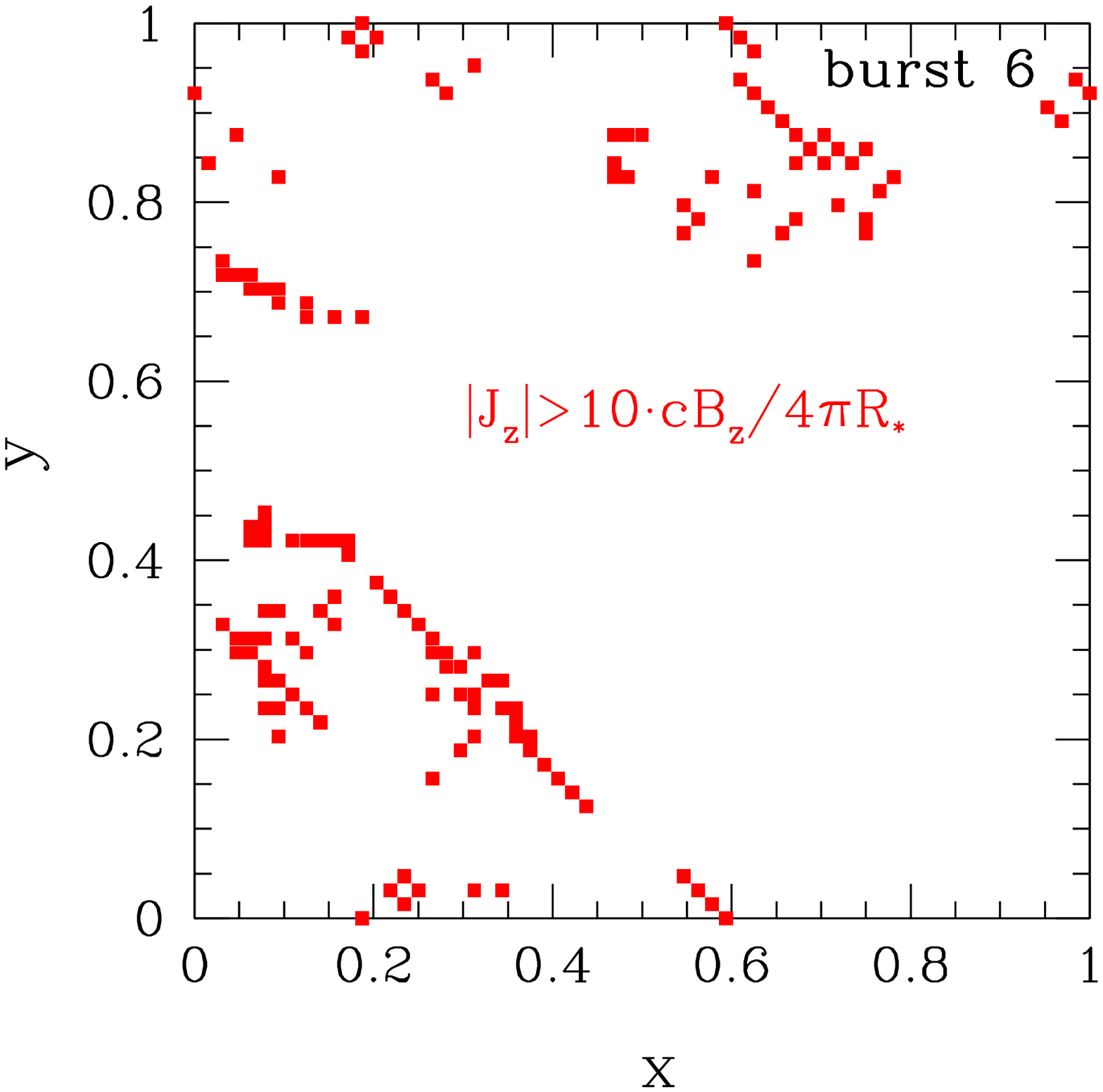}
\vskip -1.1in
\plotone{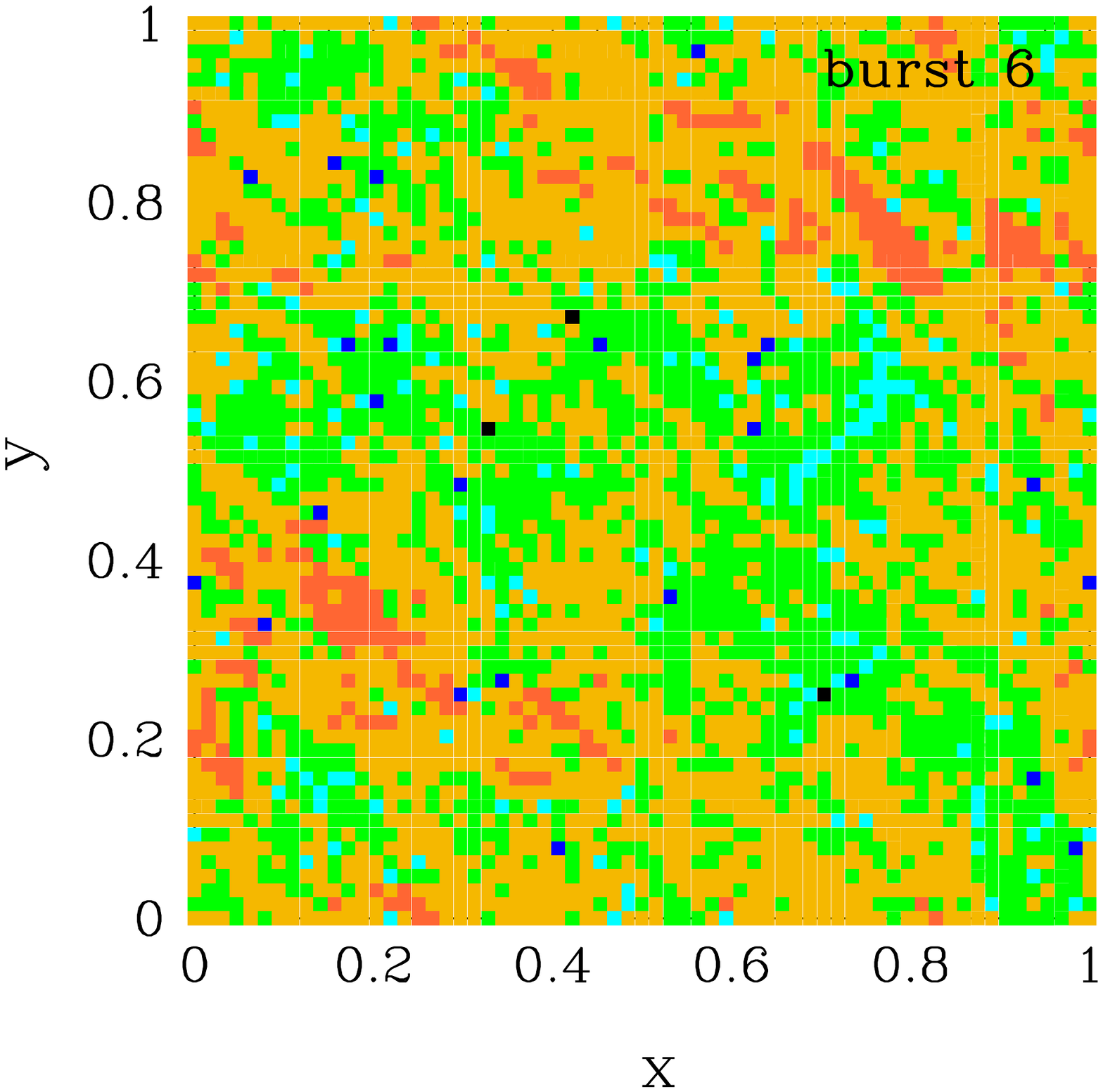}
\vskip -0.7in
\caption{Change in external magnetic energy ({\it top panel}), vertical current density ({\it middle
panel}) and internal magnetic energy ({\it bottom panel}) during the same outburst whose plastic
dissipation profile appears in Figure \ref{fig:2Danis}.  Energy scale per pixel the same
(ranging from $> 10^{39}$ up to $> 10^{43}$ erg per pixel).}
\vskip .2in
\label{fig:burst6}
\end{figure}

\begin{figure}
\vskip -0.4in
\epsscale{1.15}
\plotone{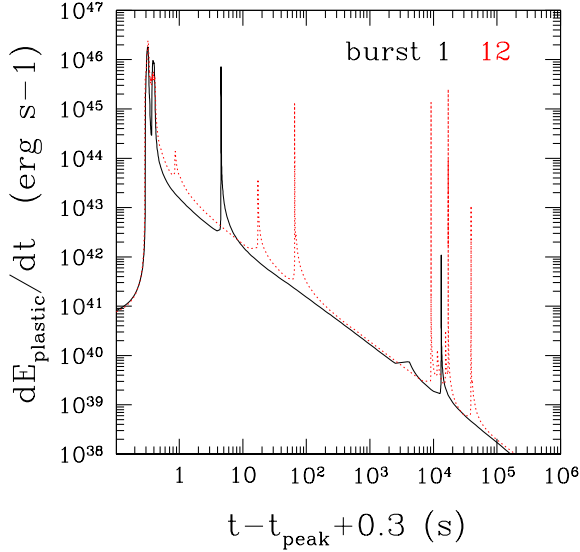}
\vskip -0.9in
\plotone{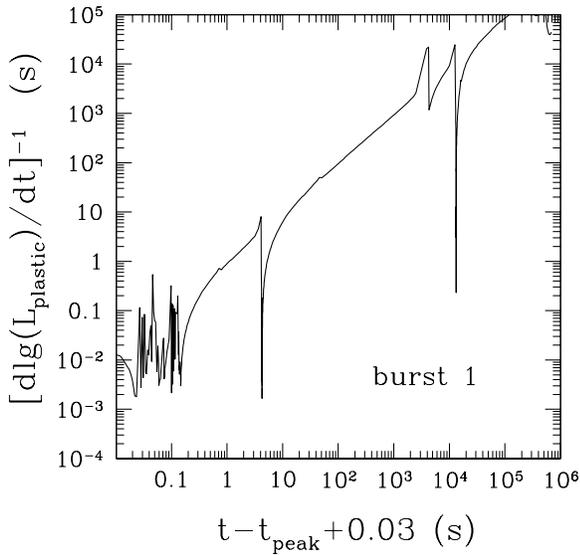}
\vskip -0.6in
\caption{{\it Top panel:}  Plastic dissipation rate versus time in the first (black line) and twelfth (dotted red line) major
outbursts seen in the 2-dimensional simulation shown in Figure \ref{fig:2Danis}. {\it Bottom panel:}  Rise and decay time of the plastic dissipation.}
\vskip .2in
\label{fig:2bursts}
\end{figure}

\begin{figure}
\vskip -0.4in
\epsscale{1.1}
\plotone{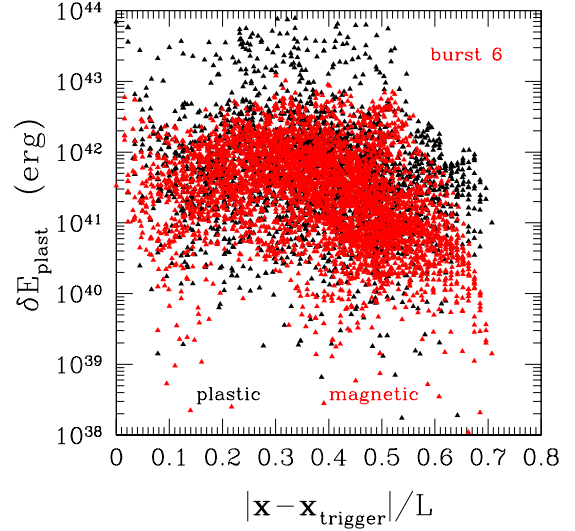}
\vskip -0.7in
\caption{Pixel-by-pixel energy release, versus horizontal distance from the 
trigger site, in burst 1 shown in Figure \ref{fig:2bursts}.  Threshold for burst onset
corresponds to creep rate $> 30$ times recent average in some pixel.  
Black points: internal plastic dissipation;  red points: external change in magnetic energy.}
\vskip .2in
\label{fig:Espread}
\end{figure}

\section{Giant Magnetar Flares}\label{s:five}

A large energy release ($\sim 10^{45}$ erg) is observed in our 2-dimensional model when runaway creep starts within a
single pixel and spreads to other locations in the crust.  This energy is partitioned between
internal plastic heating and magnetic deformation outside the star, with a compensating decrease in internal
magnetic energy.   The net deformation in a small subset of $2^{12}$ pixels can approach $\delta\varepsilon \sim 1$,
corresponding to a plastic energy release $> 3\times 10^{43}$ erg.  The total outburst energy represents a sum over
many pixels.  Indeed, the peak of dissipation repeatedly moves around, as is seen in Figure \ref{fig:2Danis}.
Some (but not all) of this movement represents repeat bursts, which are produced by the model with a range of 
energies during the $\sim 10^6$ s following the main dissipation peak.

Figure \ref{fig:Eburst} shows the various components of the energy shift for several giant-flare like events that
are drawn from the same simulation as Figure \ref{fig:2Danis}.   The change in external magnetic field is calculated
after the fact, but is not included in the time evolution equation for $\Psi_{\rm pl}$ due to the complications of implementing a realistic external magnetic geometry.   We approximate
\be\label{eq:bex}
\delta {\bf B}_{\perp,\rm ex} = {\delta\bxi \over L_{\rm mag}}B_z,
\ee
where $L_{\rm mag} = 10$ km is an effective length of the magnetospheric cavity.   On this basis, we expect that
$20$-$30\,\%$ of the change in internal magnetic energy is communicated to the exterior in the form of magnetic shear
or twist (compare the black to the cyan curve in Figure \ref{fig:Eburst}).  
Slightly more energy could be transferred to the magnetosphere if it were already twisted (see Section \ref{s:six} 
for local calculations demonstrating this).

The 2-dimensional pattern of magnetic energy injection outside the star, and the corresponding reduction in the core,
is shown in Figure \ref{fig:burst6} for the same outburst whose plastic dissipation profile appears in Figure
\ref{fig:2Danis}.  
The current flowing through the magnetar surface after the outburst is highly inhomogeneous (see the middle panel
of Figure \ref{fig:burst6}).   These zones of strong magnetic shear are a promising source of non-thermal X-ray
emission.

A fundamental point about energy conservation during magnetic field decay is raised by Figure \ref{fig:Eburst}.
One observes that the magnitude of the change in internal magnetic energy exceeds the sum of the plastically
dissipated energy and the change in elastic energy.  Some of the energy difference must be deposited in hydromagnetic
motions in the fluid core \citep{TD2001, Levin06}, which are implicitly assumed to have damped between successive
timesteps.   Core oscillations are an interesting driver of low-frequency QPO behavior, although the
spectrum has not been convincingly established. 

To see this, we refer back to the elastic equilibrium defined by Equation (\ref{eq:psiel}).   Defining
$L_\perp$ as the horizontal gradient scale for the solid stress $\sigma_{xy}$, one has
\be
|\sigma_{xy}| \sim {B_\perp B_z\over 4\pi} {L_\perp\over \delta R_c} {1\over 1+ {\cal R}_B};
\quad {\cal R}_B \equiv {B_z^2\over 4\pi \bar\mu}{L_\perp^2\over \delta R_c h_{\rm core}}
\ee
and the ratio of elastic energy in the crust to available magnetic energy in the core is 
\be
{\delta R_c\cdot \sigma_{xy}^2/2\bar\mu \over h_{\rm core}\cdot B_\perp^2/8\pi} \sim {{\cal R}_B\over (1+{\cal R}_B)^2}.
\ee
This ratio is always less than unity, and the plastic dissipation therefore does not fully compensate the change in 
core magnetic energy.

\subsection{Time Profile}

Figure \ref{fig:2bursts} shows the dissipation profiles of two outbursts (the first and twelfth) from the same calculation
as in Figure \ref{fig:2Danis}.  The duration of the main pulse ($\sim 0.1$ s) is comparable to the 
time for unbalanced stresses to redistribute around the star at the limiting creep speed $\sim 0.1\,V_\mu$. 
There is a decaying power-law tail of emission, scaling as $t^{-1}$ and with total energy $\sim 10^{-2}$ of
the main pulse energy, which is powered by continuing plastic creep.   The light curve as shown {\it does not} include emission from a trapped fireball \citep{TD1995} or from current dissipation in the magnetosphere
\citep{TLK,BT07}.  Excess decaying persistent X-ray emission with a similar power-law index ($-0.7$) and 
relative energy was observed following the 1998 giant flare of SGR 1900$+$14 \citep{woods01}.  Similar power-law
behavior is seen following shorter SGR bursts (e.g. \citealt{ibrahim01, lenters03}), but our 2-dimensional simulations
do not have the resolution to adequately address the time-profiles of these lower-energy events.

The outburst rise time is $\sim 10^{-3}$ s, which is comparable to that observed in giant flares.  There are also
repeated injections of energy near the peak of the dissipation, similar to the actual burst lightcurves recorded by
\cite{terasawa05,tanaka07}.  Our calculation represents
the accumulation and release of internal stresses.   Although magnetospheric currents
will slightly lag the internal motions, one can expect current-driven instabilities eventually to be triggered,
with slightly faster rise times over a scale of $\sim 10$ km \citep{TD1995,Lyutikov03,gill10}.  Faster
rise in the internal crustal dissipation could also be obtained by modifying the limiting creep speed to a larger
multiple of $V_\mu$ than $0.1$.

Figure \ref{fig:Espread} provides an explicit demonstration of how the energy release (both internal, black points)
and external (red points) is distributed in a non-local manner.  (The black lines in Figure \ref{fig:2Danis} only
show the shifting position of peak creep.)  In fact, the peak of energy release does not necessarily
coincide with the initial trigger point.

The total flare energy repeatably reaches $\sim 10^{45}$ erg (including changes in external magnetic energy), but
does not approach the $6\times 10^{46}$ erg that was observed from the 2004 giant flare of SGR 1900$+$14
(assuming isotropic emission: \citealt{hurley05,palmer05}).  Larger energies are possible if the magnetar core passes
through a transition to hydromagnetic instability.  Based on our model we infer that this was the case for the 2004 flare.  

Repeat bursts of intermediate energy appear in these simulations, and are also observed following giant flares
(e.g., the $\sim 4$ s burst of energy $\sim 10^{43}$ erg detected the day after the 1998 giant
flare:  \citealt{ibrahim01}).   An analysis of the energy distribution of bursts is beyond the scope
of this paper:  a proper analysis would require introducing physical scales smaller than $\delta R_c \sim 0.3$ km.
Sizable repeat bursts appear to be too common, in comparison with the limited sample of two well resolved giant
flares.  In this regard, it should be kept in mind that the moving locus of peak activity allows the burst
to probe crustal zones with differing internal magnetizations, and differing susceptibility to rapid creep.

\section{Crust-magnetosphere coupling}\label{s:six}

We now turn to consider vertical magnetic gradients within the crust.  We calculate their evolution,
starting from a state of near magnetoelastic balance and then developing runaway creep in response to an applied global stress.   The stable stratification of the stellar crust \citep{Lattimer1981,GR92} impedes vertical buoyant motion
of the magnetic field, as is observed in the early stages of Solar flares, but still allows a coupling to the magnetosphere
via sideways shear motions.  We find that magnetic shear or twist can be ejected into the magnetosphere from deep in the crust in a few milliseconds, much shorter than the durations of most SGR bursts.  

We have described the crust-core Maxwell stress by an incompressible 2-dimensional horizontal shuffling of vertical
field lines.   The rigidity of the crust prevents the displacement of the external field lines 
by plastic creep (Equation (\ref{eq:bex})) from instantly mirroring the core displacement.  We can compare the
gradient energy that accumulates in the crustal magnetic field with the energy that would be stored in the 
magnetosphere {\it if} the shuffling of the external field lines were smooth and instantaneous (that is, if
effectively the crust were absent).  The answer is sensitive to the presence of a background magnetic shear
$\delta B_{\perp,0} \sim (\xi_{\perp,0}/R_\star)B_z$ associated with a radial current flowing across the magnetar
surface \citep{TLK}.  The additional crustal gradient energy associated with a horizontal displacement $\xi_\perp$ is
\ba\label{eq:delE}
\delta E_B &\sim& \delta R_c \left[{(\delta B_\perp^c)^2\over 8\pi} + 
                {\delta B_\perp^c \delta B_{\perp,0}\over 4\pi}\right]\nn
&\sim& R_\star {B_z^2\over 8\pi} \left[{R_\star\over \delta R_c}\left({\xi_\perp\over R_\star}\right)^2 + 2
{\xi_\perp \xi_{\perp,0}\over R_\star^2}\right]
\ea
per unit area, where $B_\perp^c \sim (\xi_\perp /\delta R_c)B_z$.   If the same displacement field
were immediately to spread to the magnetosphere, then the factor of $R_\star/\delta R_c$ in the first 
term on the right-hand side of the estimate (\ref{eq:delE}) would effectively be absent.   

The net result is that trapping of vertical magnetic shear in the crust will enhance the magnetic shear energy
by a factor $\sim R_\star/\delta R_c$ if background shear is absent;  but only provides 
an order-unity enhancement if the background shear is relatively strongly, $|\xi_\perp| \lesssim |\xi_{\perp,0}|$. 
In this section, we consider how the magnetic shear relaxes to a smoother, lower energy state.

In contrast with the 0-dimensional model described in Section \ref{s:two}, there is no interference between the
local stress and the applied global stress, because they are orthogonal.   The vertical gradient of $B_i B_z/4\pi$ 
within the crust (here $i = x,y$) is compensated by a gradient in $\sigma_{iz}$, and the total stress is 
\be\label{eq:sigtot}
\sigma = \sqrt{(\sigma_g)_{xy}^2 + (B_xB_z/4\pi)^2 + (B_yB_z/4\pi)^2}.
\ee
Here we write explicitly the dominant component of the global stress.  

This situation is also distinguished from the spot and fault models by the strong vertical inhomogeneity of the
shear modulus and the yield stress.  The question therefore arises as to whether temperature feedback is needed
to trigger creep runaway.  Here we show that it is not: a small displacement of the transverse magnetic field upward
in the crust increases the ratio of Maxwell stress to shear modulus and causes in exponential growth of the creep rate.

The vertical magnetic shear begins to evolve rapidly as $\sigma_g$ builds up and
the creep time in the lower crust drops below the growth time of the applied stress.  
It may appear at first sight that the two Maxwell stress terms are subdominant in the right-hand side of Equation
(\ref{eq:sigtot}) when $\sigma_{xy}$ approaches the yield stress, because $\sigma_{xy} \sim (R_\star/\delta R_c)
(B_\perp B_z/4\pi)_-$. But a stronger embedded horizontal magnetic field is expected to develop in plastic fault zones
which have width comparable to depth (crustal thickness $\delta R_c$), and a net displacement $> \delta R_c$
along the fault.   This suggests that vertical energy leakage is localized in the yielding zones,
which we show Section \ref{s:eight} has interesting implications for magnetar QPO activity.

\subsection{One-dimensional Vertical Evolution}

\begin{figure*}
\includegraphics[width=1.0\textwidth]{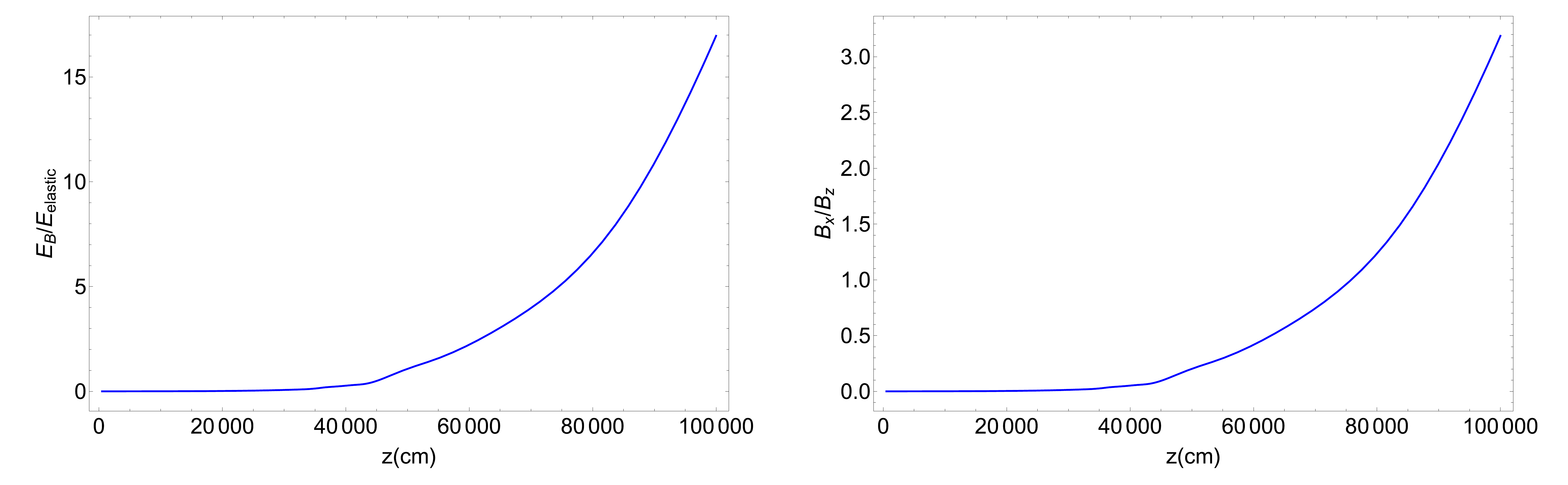}
\caption{Vertical profile of a neutron star crust that starts in a slowly deforming magnetoelastic
state:  creep rate $\dot{\varepsilon}=0.003$ s$^{-1}$ (independent of depth) and $B_z = 10^{15}$ G.  
{\it Right panel:}  Ratio of transverse to vertical magnetic field $B_x/B_z$ is
obtained from the steady solution Equation (\ref{eqwave2}) using the stress formula (\ref{eqy}).
{\it Left panel:}  Available magnetic energy compared with elastic energy, as a function of depth $z$.
Crust model described in the text.}
\vskip .2in
\label{fig:bac}
\end{figure*}

We consider a vertical 1-dimensional test problem which illustrate the dynamical ejection of magnetic shear.
The crust is threaded by a uniform vertical magnetic field $B_z = 10^{15}$ G that is bent by a variable amount in the 
horizontal $x$-direction, $B_x(z) = \partial_z\xi_{B,x} B_z$.  In the initial condition, the Maxwell stress $B_x B_z/4 \pi$ 
saturates the yield stress $\sigma_b$ (Equation (\ref{eqy})) at all depths $z$ below the magnetar surface,
and the creep rate is uniform.  The corresponding transverse field, shown in Figure~\ref{fig:bac}, equals $B_z$ 
only at a particular depth.
Everywhere else $B^2/4\pi = [(B_x/B_z) + (B_z/B_x)]\sigma_b > 2\sigma_b$, meaning that the field needed
to induce plastic creep is generally stronger than $(4\pi\sigma_b)^{1/2}$ in the absence of an addition global stress.

The vertical dependence of 
crustal density, melting temperature, and shear modulus used in this paper are constructed by fitting the
data from nuclear equation of state calculations, both below \citep{Negele1973} and above \citep{Haensel1989} the neutron
drip point, for a surface gravity $g = 2 \times 10^{14}$ cm s$^{-2}$.  Smooth fitting functions are applied to the density
and shear modulus profiles $\rho(z)$, $\mu(z)$, which allows us to avoid complications associated with discrete jumps
in these variables.

Figure \ref{fig:bac} also shows the ratio between $B^2_x/(8\pi)$ and the initial elastic energy density 
${1\over 2}\sigma_b^2/\mu$, as a function of depth in the crust.  The magnetic field is the 
dominant energy reservoir near the base of the crust, where most of the potential energy is concentrated.  

\begin{figure}[t]
\plotone{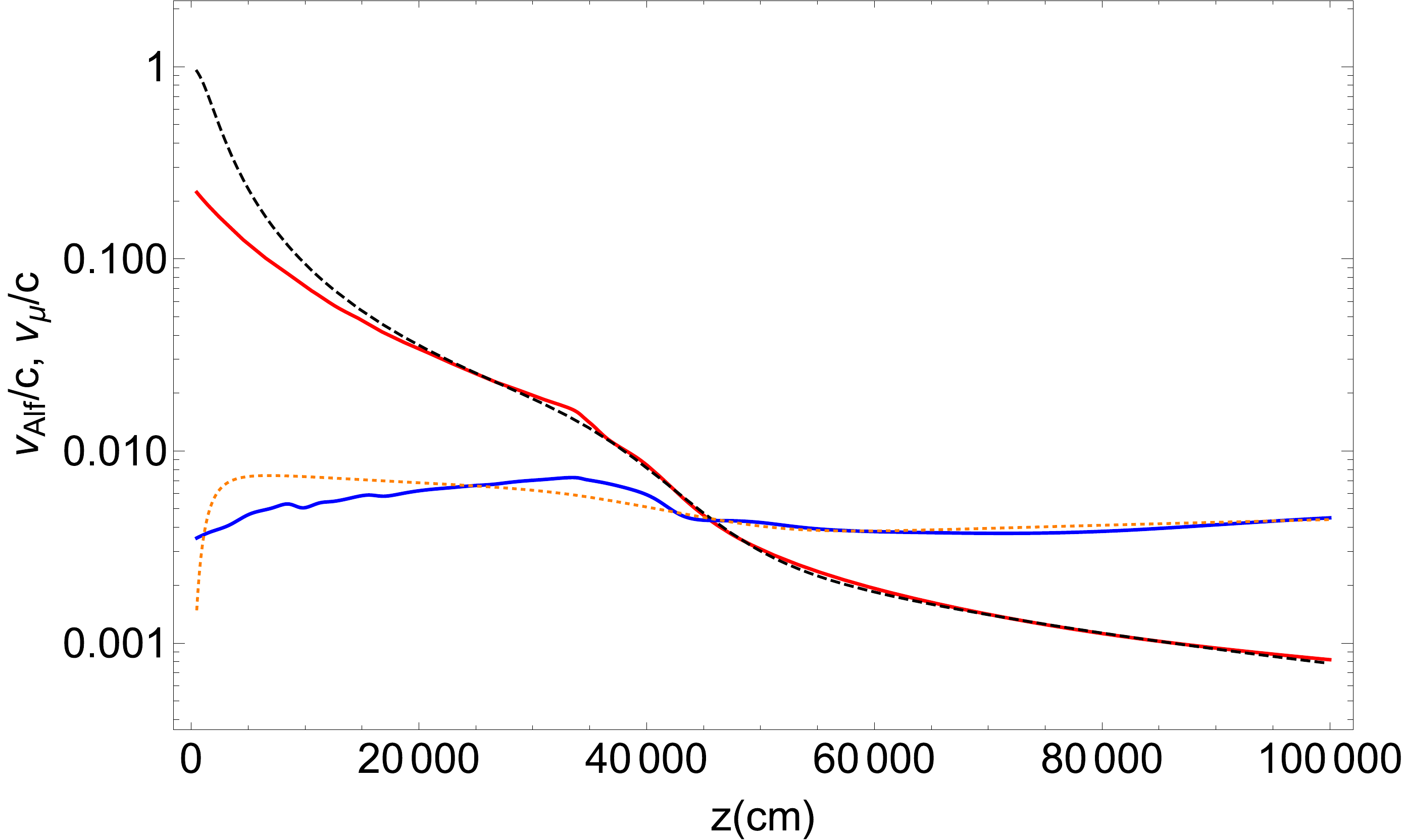}
  \caption{Vertical  profile of Alfv\'en speed $V_{\rm A}$ (red, thick line) and its fitting curve (black, dashed line);
and of transverse shear wave speed $V_\mu$ (blue, thick line) and its fitting curve (orange, dotted line). 
In the upper part of the crust $V_\mu \ll V_{\rm A} \simeq c$.}
\vskip .2in
\label{fig:rho}
\end{figure}

\begin{figure*}
\includegraphics[width=1.0\textwidth]{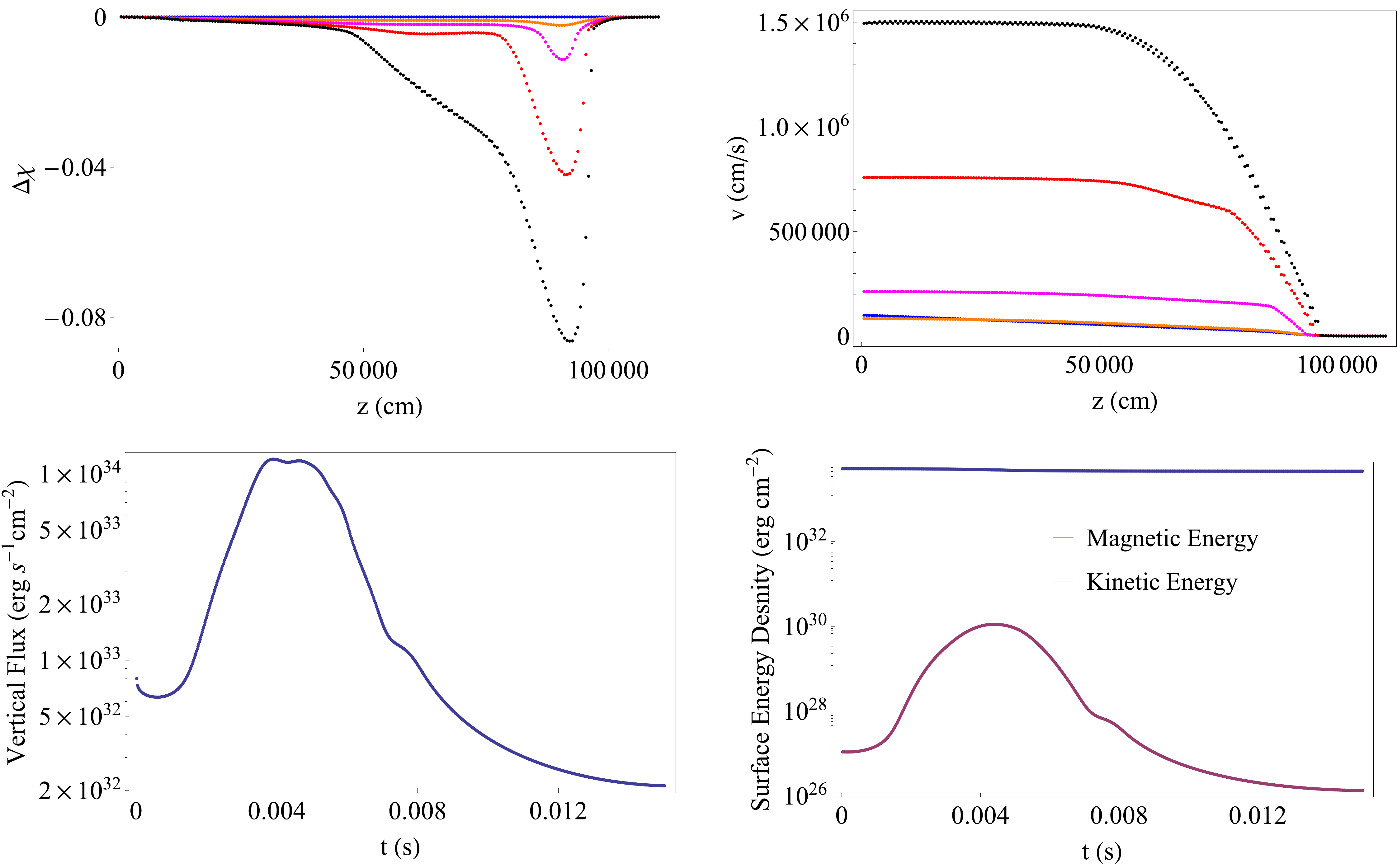}
\caption{{\it Top panels:}  Evolution of $v$ (left) and $\chi-\chi |_{t=0}$ (right) at time $0$ ms (blue), $1$ ms (orange), $2$ ms (magenta), $3$ ms (red) and $4$ ms (black). 
{\it Bottom panels:}  Evolution of vertical Poynting flux measured at the star surface, and integrated surface 
energy density (kinetic $+$ magnetic) during the first 15 ms.  The maximum in kinetic energy roughly coincides 
with the maximum of outgoing energy flux, both being driven by unwinding of the initial magnetic shear.}
\vskip .2in
\label{fig:gpplot}
\end{figure*}

The evolution away from the initial slow creep state can be encapsulated in two variables
\be
v \equiv \partial_t \xi_{B,x},  \quad \chi \equiv \partial_z \xi_{B,x}.
\ee
Freezing of magnetic flux in the plastically deforming crust implies that
\be\label{eqwave1}
\partial_t \chi = \partial_z v,
\ee
so that $v$ is the combined transverse velocity of magnetic field and crustal material.
The transverse component of the relativistic Euler equation reads
\be\label{eqwave2}
 \partial_t v =  V^2_{\rm A} \partial_z \chi + \frac{1}{\rho_B} \partial_z \sigma_{xz}.
\ee
The magnetic field dominates the effective inertia in the upper crust, 
$\rho_B=\rho+B^2_z/4 \pi c^2$, so that the Alfv\'en speed $V_{\rm A} = B_z/(4 \pi \rho_B)^{1/2}$
is limited to the speed of light $c$.  Figure~\ref{fig:rho} shows the vertical dependence 
of $V_{\rm A}$ and transverse shear wave speed $V_\mu =\sqrt{\mu/\rho_B}$.
The initial equilibrium state is
\be\label{eqi1}
\left (\chi+ {4 \pi\over B^2_z} \sigma_{xz} \right )_{t=0}=\chi_0,
\ee 
where $\chi_0 \sim \mathcal{O}(0.1)$ is the constant magnetic shear that is imposed by the magnetosphere.

Because we are studying the transition between magnetoelastic equilibrium and a dynamical state of the crust,
we cannot assume that at each timestep the crust relaxes to magnetoelastic balance.  Therefore we 
approximate $\partial_z v = \dot{\varepsilon}_{\rm pl}$, and use Equation (\ref{eqy}) to write
\be\label{eqsxz1}
\sigma_{xz} = \sigma_b(T, \partial_z v) {\rm sgn}(\partial_z v).
\ee
The yield stress is insensitive to small changes in $\dot{\varepsilon}_{\rm pl}$ following Equation (\ref{eqy}).

Near the top of the crust, $\mu$ is relatively small and the term involving $\sigma_{xz}$ is 
negligible in Equation (\ref{eqwave2}).  In this regime the crustal material behaves like a magnetofluid
($V_{\rm A} \sim c \gg V_{\mu}$):  the deformation is driven mainly by the Maxwell stress, and plastic
flow simply follows the magnetic field.  In some circumstances the upper parts of the magnetar crust may
actually be in a fluid state; but the melt depth of a quiescent magnetar is typically shallow enough that $V_{\rm A} \simeq c$,
and the solid-to-fluid transition has a small influence on the dynamics.

We impose an outgoing wave boundary condition at the upper boundary, 
 \be\label{eqb1}
(\partial_t-V_{\rm A} \partial_z) v\,\Bigr|_{z=0}=0.
\ee
In the deeper part of the crust $V_{\rm A} < V_\mu$.   We keep the evolution
short enough that the initial magnetoelastic equilibrium is only slightly disturbed near the lower boundary,
and therefore simply assume that the magnetic field is pinned there with constant Maxwell stress,
\be\label{eqb2}
v(z=\delta R_c,t) =0; \quad \chi(z=\delta R_c,t)={\rm const}.
\ee
To avoid numerical instabilities, the uniform creep rate $\dot\varepsilon_{\rm pl}$ is smoothly set to zero
in a thin layer near the lower boundary.   It is also essential to smoothly interpolate the ${\rm sgn}$
function in Equation (\ref{eqsxz1}), which otherwise is discontinuous when $\partial_z v$ changes sign:
\be\label{eqsm}\label{eq:smooth}
 {\rm sgn}(\partial_z v) \rightarrow  \frac{\partial_z v}{\sqrt{(\partial_z v)^2+\dot{\varepsilon}_0^2}},
\ee
Here $\dot\varepsilon_0$ is a small parameter characterizing the thickness of the transition.
With this modification, Equation (\ref{eqwave2}) takes a form similar to heat diffusion equation 
with spatially variable thermal conduction coefficient.

Next an external stress $\sigma_{xy}$ is smoothly added to the system, representing (e.g.) a horizontally 
propagating elastic wave.   This means that, at a fixed vertical creep rate, the vertical stress is {\it reduced} to
\be\label{eqnews}
\sigma_{xz} = {\rm sgn}(\partial_z v) \sqrt{[\sigma_b(T,\partial_z v)]^2-\sigma^2_{xy}}.
\ee
Strictly speaking, $\dot{\varepsilon}_{\rm pl}^2 =(\partial_z v)^2+\dot{\varepsilon}_{xy}^2$, where the
second term is the horizontal strain rate.  As the timescale for horizontal stress growth is 
longer by a factor $\sim R_\star/\delta R_c$, we approximate $\dot{\varepsilon}_{\rm pl} 
\approx |\partial_z v|$ during the 1-dimensional evolution.  

Collecting Equations (\ref{eqwave1})-(\ref{eqnews}), we obtain a complete 
system to model the wave-induced stress imbalance and vertical hydromagnetic wave ejection.

\subsection{Numerical Results}

Figure~\ref{fig:gpplot} shows the result of introducing an external `driver' wave stress 
\be
\sigma_{xy} = \varepsilon_{xy} f(z) \sin(\omega t)\,\mu.
\ee
The radial eigenfunction $f(z)$ reproduces the ${}_2t_0$ global elastic mode \citep{McDermott1988},
which has a frequency $\omega \simeq 2^{1/2} \bar V_\mu/R_\star$.  
To obtain numerically clean results, we take a driver amplitude $\varepsilon_{xy} = 0.06$.
The background creep rate is $\varepsilon_{\rm pl} = 1$ s$^{-1}$, which is significantly slower
than the ensuing evolution.   The smoothing parameter $\dot\varepsilon_0$ in the stress formulae (\ref{eq:smooth}),
(\ref{eqnews}) is necessarily lower than this, and is taken to be $\sim 0.1$ s$^{-1}$, as limited by numerical precision.

The net result is a burst of upward Poynting flux $S_P$ that builds up significantly over a few milliseconds, tapping
the crustal magnetic shear $\chi$.  The Poynting flux is proportional to $\chi$,
\be
S_P = \frac{B^2_z}{4 \pi} \chi v,
\ee
and is significantly enhanced in the upper crust if the shear has a baseline value $\chi_0$ that imposed by
external magnetic twist or shear.  The net electromagnetic energy ejected to the magnetosphere is
\be
\int dt\,S_P  \approx \frac{B^2_z}{4 \pi}\,\chi_0\, \Delta x|_{z=0},
\ee
where $\Delta x$ is the net horizontal displacement of the field lines.  
Taking $\chi_0=0.1$, this works out to about $7\%$ of the total energy loss over the first 10 ms of
the evolution.  The majority of the change in magnetic gradient energy is dissipated within the crust, 
at a rate $\sigma_b |\dot{\varepsilon}|$ per unit volume.  Numerical tests show that this efficiency is insensitive
to the amplitude of the driver elastic wave.

The magnetic tilt fluctuation is small near the stellar surface, $\chi \simeq v/V_A \sim v/c$, and consequently the
plastic creep rate is also suppressed, $\dot{\varepsilon} \sim \chi/(10~{\rm ms}) \sim v/(c~\cdot~10~{\rm ms})$.
Combining this with the relatively small shear modulus, we find negligible plastic heating at shallow depths.   The
crust may behave like a fluid even while it is still in the solid phase.  

What happens to the magnetic shear after it is ejected from the crust depends on details such as the length of
the magnetospheric flux bundle, and the horizontal twist structure which is not captured by this 
1-dimensional model.   Effectively we have
taken an infinitely long magnetospheric cavity, so that reflection of the outgoing `wave' can be neglected.  
This corresponds in a dipole geometry to a field line extending to a maximum radius $R_{\rm max} \sim {1\over 3}
c \cdot 3~{\rm ms} \sim 30\,R_\star$.  Here we make the implicit assumption that twist ejection is accompanied
by enough pair creation or particle ejection to sustain the MHD approximation in the magnetosphere.   Then
the twist propagates with a speed $V_A \sim c$ over a cavity of length $\sim 3R_{\rm max}$.
In more compact parts of the magnetosphere, a quasi-static but strongly
localized twist is established, which probably is subject to secondary current-driven instabilities.   Details
of the subsequent damping are not investigated here.

\begin{figure}
\vskip -0.3in
\epsscale{1.15}
\plotone{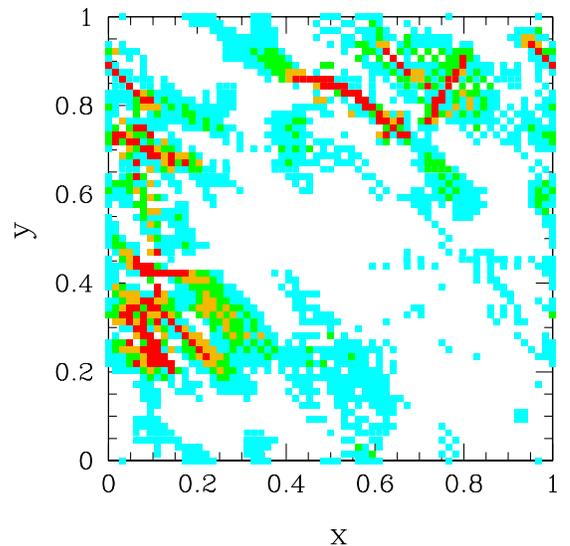}
\vskip -0.7in
\caption{Temperature profile right after the peak of the outburst whose dissipation profile is shown in
Figure \ref{fig:2Danis}.  Pixel colors red, gold, green and cyan correspond to $T > 0.5$, $0.3-0.5$, $0.2-0.3$, and
$0.1-0.2\,T_{\rm melt}$ at density $10^{14}$ g cm$^{-3}$.}
\vskip .2in
\label{fig:2Dtemp}
\end{figure}

\begin{figure*}
\includegraphics[width=1.0\textwidth]{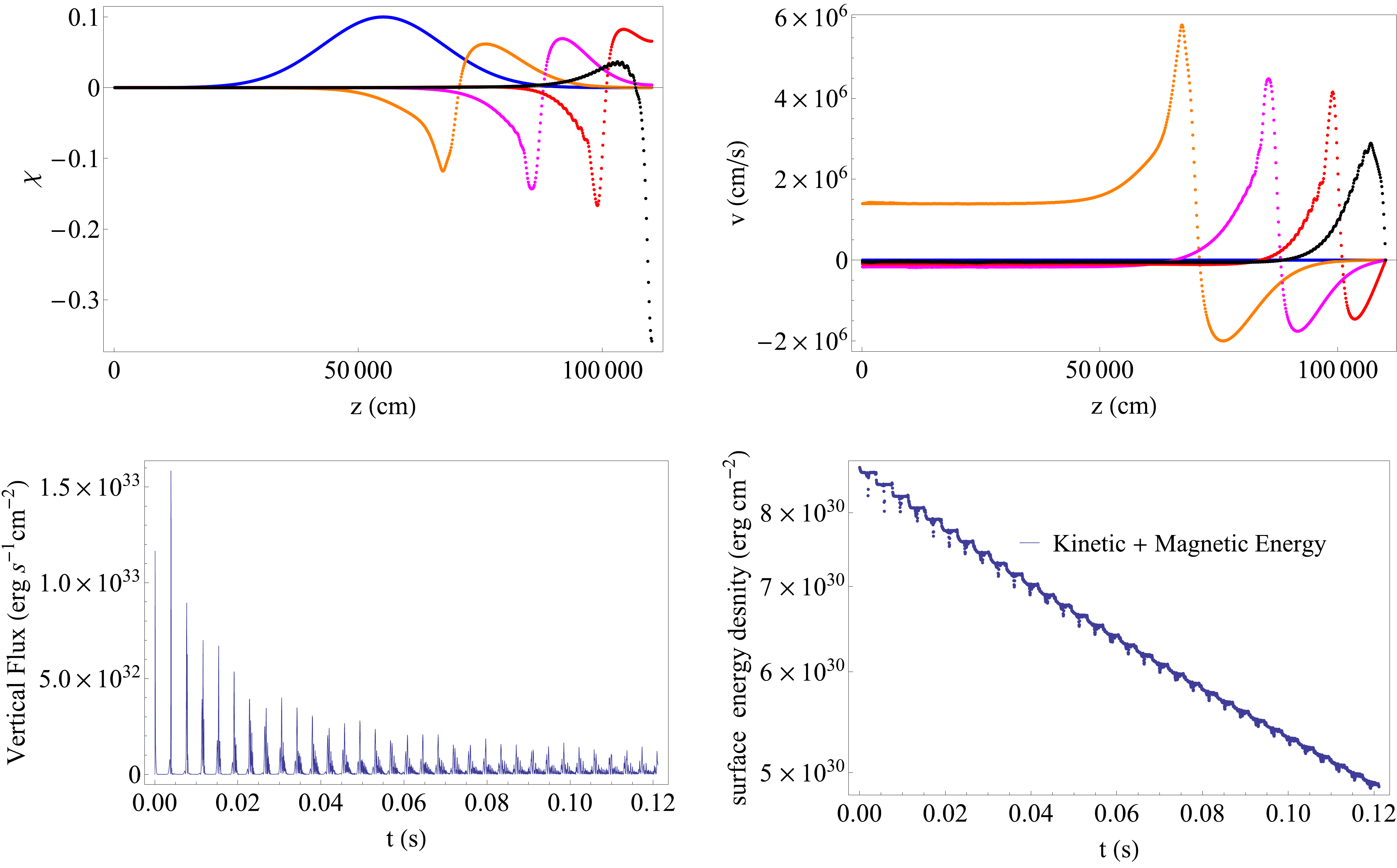}
\caption{Evolution of twist and kinetic energy in a melted crust, showing decaying periodic behavior.  
{\it Top panels:}  Evolution of $v$ (left) and $\chi$ (right) at time $0$ ms (blue), $0.5$ ms (orange), $1$ ms (magenta),
$1.5$ ms (red) and $2$ ms (black).  The initial profile of $\chi$ is given by Equation~(\ref{eqinitchi}) and initially 
$v = 0$.  {\it Bottom panels:}  Evolution of vertical Poynting flux measured at the star surface, and integrated surface
energy density (kinetic $+$ magnetic) during the first 0.12 s.  The time-integrated Poynting flux balances the loss of
magnetic shear energy from the resonant cavity.}
\vskip .2in
\label{fig:fluidplot}
\end{figure*}

\section{Trapped Alfv\'en Waves in Liquified Patches}\label{s:seven}

Narrow zones of strong plastic flow experience melting
in our 2-dimensional simulations.  Figure \ref{fig:2Dtemp} shows the temperature profile just after
the peak of the outburst shown in Figure \ref{fig:2Danis}.  
The energy needed per unit volume to melt the crust started at a temperature $\ll T_{\rm melt}$ is
\be
\Delta U  = \int^{T_{\rm melt}} dT C_V(T) \sim 0.08\mu.
\ee
(Here we include the specific heat of the ions as calculated by \cite{chabrier93}, and assume
that the specific heat of dripped neutrons is suppressed by superfluidity.)  
This compares with an elastic energy at the yielding point 
\be
{1\over 2}\varepsilon_{\rm el}^2\mu \sim 
\left\{\begin{array}{rcl}
                  1\times 10^{-4}\mu & \quad (T \sim T_{\rm melt})\\ 
                  1\times 10^{-2}\mu & \quad (T \sim 0.02\,T_{\rm melt}).
\end{array}\right.
\ee
Thus, when the crust starts relatively cold, it must continue to be deformed through $\sim 5$-$10$ times 
the yield strain in order to melt, increasing to $\sim 30$-$10^2$ times when it starts close to melting.   
Such a large relative displacement is possible if the solid stress is sourced by distant Maxwell stresses,
and occurs in the pixels in Figure \ref{fig:Espread} with dissipated energy $\gtrsim 10^{43}$ erg.

Let us now consider the evolution of an inhomogeneous magnetic shear in such a melted layer.  The 
corresponding time evolution equations are Equation (\ref{eqwave1}) and (\ref{eqwave2}) with $\sigma_{xz}$
set to zero.   Now the magnetic field provides the only restoring force.

The boundary condition Equation~(\ref{eqb1}) is still valid, and at $z=\delta R_c$ we continue to impose $v=0$. 
To study the ejection of a trapped Alfv\'en wave, we prepare an initial data such that $v(z,t=0)=0$ and
\be\label{eqinitchi}
\chi(z,t=0)= 0.1 e^{-40(z-\delta R_c/2)^2/d^2}.
\ee
Any constant background twist $\chi_0$ can be added to $\chi$ to obtain a new solution. In Figure~\ref{fig:fluidplot},
we show the numerical evolution using the above initial data.  Within the first $0.12$ s, the Alfv\'en wave packet
bounces within the crustal cavity for $N \sim 30$ cycles.   The quality factor extracted from the evolution of the total
energy is
\be
Q = \frac{2 N}{\log(E_N/E_0)} \sim 110.
\ee
Notice that the energy loss shown in the bottom right panel of Figure~\ref{fig:fluidplot} is consistent with integrating
the flux in the bottom left panel.  Such a high quality factor is due to the vertical stratification of fluid density.  The
density scale height $l_\rho$ decreases toward the magnetar surface ($z =0$), providing a turning point for an upward-propagating Alfv\'en wave where $k l_\rho = \omega l_\rho/V_A \propto \rho l_\rho \sim 1$.  A similar effect
is seen in the tunneling of an ideal elastic wave out of a magnetized neutron star crust \citep{Blaes1989,Link14}. 

This trapped Alfv\'en wave, being slowly damped and strongly periodic, is a good candidate for the mode 
underlying high-frequency QPO behavior.   The frequency of such a standing wave will 
exceed $\sim 1$ kHz if it has has radial nodes, or if the lower part of the crust remains solid, thereby reducing 
the Alfv\'en crossing time of the resonant cavity.

A background twist $\chi_0$ outside the star modifies the outgoing Poynting flux in the following way.  The first
half oscillation inside the star supplies an additional component of the time integral $S_P dt$.   Thereafter,
the contribution from the term in $S_P$ proportional to $v \chi_0$ averages nearly to zero.

\section{Overstability of Global Elastic Modes:  \\ Interaction with Plastic Patches}\label{s:eight}

We now consider how a global elastic mode interacts with plastically deforming zones
in the magnetar crust, and explain how it may gain energy from them.   We previously showed
that a static global stress enhances the creep rate in localized plastic `spots' and `faults' (Section \ref{s:two}).
A growing global stress triggers a localized flow of Poynting flux into the magnetosphere (Section \ref{s:six}).  
An oscillating elastic mode will produce a periodic modulation of this dissipation; and at the same
time, it can feed off the collective shear or vortical motion of the plastic zones.

\begin{figure}[t]
\epsscale{1.2}
\plotone{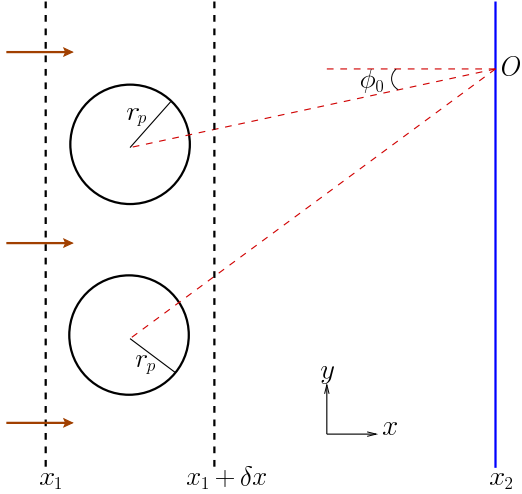}
\vskip .2in
  \caption{Global elastic response of the crust to enhanced yielding in localized plastic zones.
Secondary waves are superposed on the original wave, and their cumulative effect is measured at 
a displaced position ``O''.  This introduces a damping/growth term into the elastic wave
dispersion relation, Equations (\ref{eq:globwave}) and (\ref{eq:gr}).}
\vskip .2in
\label{fig:wave}
\end{figure}

To start, we idealize each plastic patch as a circular spot of characteristic radius $r_p$,
which we may define as the radius of maximal stress and creep rate.  Each spot is
much smaller than the wavelength of an imposed elastic wave, e.g. $r_p \ll R_\star$ (Figure~\ref{fig:wave}).
Creep within a spot is driven at a rate $\dot\epsilon_{\rm pl,0}$ 
by a Maxwell stress $(B_\phi B_z/4\pi)_-$ at the lower crust boundary.
Outside the spot the strain field is source free and satisfies the wave equation
\be\label{eq:spoteq}
{\partial^2\xi_\phi\over\partial t^2} 
= {\bar V_\mu^2\over \varpi^2} {\partial\over\partial\varpi}\left(\varpi^2 {\sigma_{\phi\varpi}\over\bar\mu}\right).
\ee
We continue to work with a vertically averged strain field, and here adopt
cylindrical coordinates centered on the spot.   Then 
$\sigma_{\phi\varpi} = \bar\mu \varpi \partial_\varpi (\xi_\phi/\varpi)$.  

The plastic flow within a spot is vortical, and at constant stress the elastic deformation outside
the spot is independent of time, satisfying $\varpi^2 \sigma_{\phi\varpi} =$ const and $\xi_\phi \propto 1/\varpi$.
But when the stress is increasing or decreasing, the elastic solid experiences a net circulation
\be
\Gamma = 2\pi r_p v_p \sim 2\pi r_p^2 {\partial_t\sigma_{\phi\varpi}(r_p)\over \bar\mu}.
\ee
An explicit construction showing this is given in the Appendix.   

The feedback of many spots on the global strain field now follows easily.  Given a spot
surface number density $n_p$ and filling factor $f_p \sim n_p \pi r^2_p$, one has
\be\label{eq:spotvor}
(\bnabla\times\dot{\bxi})_z = n_p \Gamma = 2f_p {\partial_t\sigma_{\phi\varpi}(r_p)\over \bar\mu}.
\ee

It is essential here to include the response of the Maxwell stress to accelerated creep associated with
an imposed wave stress.  Consider a planar wave strain field $\xi^y(x-\bar V_\mu t)$.  Then 
we write 
\be
{\partial_t\sigma_{\phi\varpi}(r_p)\over \bar\mu} = K_\sigma \dot\varepsilon_{\rm pl,0} {\partial\xi_y\over\partial x},
\ee
so that  Equation (\ref{eq:spotvor}) becomes
\be
{\partial^2\xi_y\over\partial x\partial t} = 2K_\sigma f_p \dot\varepsilon_{\rm pl,0}{\partial\xi_y\over\partial x}.
\ee
This represents negative damping of the strain field, with growth rate
\be
-\gamma = 2K_\sigma f_p \dot\varepsilon_{\rm pl,0}.
\ee  

To evaluate the coefficient $K_\sigma$, we need to consider the response of the 
plastic flow to the additional stress $\sigma_w$ of the elastic wave.  The creep rate (\ref{eqy1}) rises to
\be
\frac{\dot{\varepsilon_{\rm pl}}}{\dot{\varepsilon}_{\rm pl,0}} = e^{0.183 \sigma_w \bar{N} \Gamma /\mu}.
\ee 
The change in stress in the plastic spot depends crucially on the sign of the derivative
$\partial\sigma_{\phi\varpi}/\partial\varepsilon_{\rm pl}$, where $\varepsilon_{\rm pl}$ is the cumulative plastic strain:
\be
\partial_t\sigma_{\phi\varpi}(r_p) = {\partial\sigma_{\phi\varpi}\over\partial \varepsilon_{\rm pl}}\dot\varepsilon_{\rm pl}.
\ee
A positive sign corresponds to a locally unstable magneto-elastic equilibrium.
In the linear regime, we therefore recover
\be
K_\sigma \simeq 0.183 \bar{N}\Gamma \cdot {1\over\bar\mu}{\partial\sigma_{\phi\varpi}\over\partial \varepsilon_{\rm pl}}.
\ee

The same result is obtained by a direct construction.   We introduce a damping term into the wave equation
for the global elastic mode, 
\be\label{eq:globwave}
{\partial^2 v_y\over \partial t^2} + \gamma{\partial v_y\over\partial t} = \bar V_\mu^2{\partial^2 v_y\over\partial x^2},
\ee
which has the solution $v_y = v_y^+(x-\bar V_\mu t) + v_y^-(x+\bar V_\mu t)$ in the absence of damping.   
Treating the damping term as a perturbation, we consider the change $\delta v_y$ in the velocity field 
at some position $x_2$ due to spot rotation within a zone of thickness $\delta x$ near position $x_1$.  Then $\delta v_y$
satisfies
\be
-4 \bar V_\mu^2 {\partial^2\delta v_y\over\partial x_+\partial x_-} = - \gamma{\partial v_y^+\over\partial t},
\ee
where $x_\pm = x\pm V_\mu t$.   Integrating once with respect to $x_+$ and noting that $\delta x_- = 2\delta x$
at constant $x_+$, we obtain
\be\label{eq:dvy}
\delta v_y^+(x_2,t) = - {\gamma \delta x\over 2 \bar V_\mu} v_y^+(x_2) = 
-{\gamma \delta x\over 2}{\partial\xi_y^+\over\partial x}(x_2).
\ee

The velocity perturbation is obtained by linear superposiition.   The imposed elastic wave is assumed to be
periodic with frequency $\omega$ and wavenumber $k = \omega/\bar V_\mu$.  The velocity
response around each spot has the same frequency.  Hence the solution to the wave equation (\ref{eq:spoteq}) outside 
a single spot is 
\be
v(\varpi) = v_0 iH_1^{(1)}(k\varpi).
\ee
The Hankel function $iH_1^{(1)}(x) \simeq 2/\pi x$ at small argument; hence $v_0 \simeq k\Gamma/4$.

Following Figure~\ref{fig:wave}, the collective contribution to $\delta v_y$ at $x_2$ 
from all the spots between $x_1$ and $x_1 + \delta x$ (denote $\Delta x \equiv x_2-x_1$) is
\ba
\delta v_y &=& n_p \delta x \int^{\pi/2}_{-\pi/2} d \phi_0 v_0\,iH_1[k \Delta x \sec \phi_0] \Delta x \sec \phi_0
\nonumber\\
&=& 2{v_0\over k}n_p\delta x e^{ik\Delta x}. %
\ea
Identifying this with Equation (\ref{eq:dvy}), and substituting for $v_0$, we find growth at a rate
\be\label{eq:gr}
-\gamma = {n_p\Gamma \over \partial\xi_y/\partial x} = 2K_\sigma f_p\dot\varepsilon_{\rm pl,0}.
\ee

We conclude that a global elastic mode can experience ``super-radiant'' amplification by 
scattering off localized plastic spots in the magnetar crust, {\it if} a localized Maxwell stress
driving creep in these spots {\it increases} with displacement from the initial equilibrium.
In other words, parts of the magnetar crust must be in a hydromagnetically unstable state.   

The dissipation rate in these plastic zones is modulated by the global mode, which feeds off
part of the magnetic energy liberated.   Some of these energy is transported into the magnetosphere
(Section \ref{s:six}).   The power output from the boundary of a single spot is
\be
\frac{dE_p}{dt} \sim  2 \pi r_p\cdot \dot\varepsilon_{\rm pl} r_p \cdot \sigma_w,
\ee
and the power output per unit area is (with $E_w \sim \sigma_w^2/\bar\mu$ and
$\mu^{-1}\partial\sigma_{\phi\varpi}/\partial\varepsilon_{\rm pl} =$ O(1))
\be
\frac{d E_w}{d t} \sim n_p \frac{dE_p}{d t} \sim 0.183 \bar{N} \Gamma\cdot 2f_p \dot{\varepsilon}_{\rm pl,0}\, E_w
\sim \gamma E_w.
\ee
The corresponding energy growth per cycle $r \sim \gamma P = \gamma 2 \pi/\omega$ is
\be
r \sim 0.01 \left (\frac{\dot{\varepsilon}_{\rm pl,0}}{0.0025} \right ) \left (\frac{f_p}{0.1} \right ) \left ( \frac{T}{0.5 T_m}\right )^{-1} \left ( \frac{P}{30 {\rm ms}} \right ).
\ee

Temperature feedback may have a negative feedback on super-radiant amplification of elastic waves,
because it implies a decrease in the equilibrium Maxwell stress that can be sustained inside a plastic
spot.  

We have found that the geometry of the plastic zones can have a strong impact on the feedback process.
One can show that an extended and connected (e.g. equatorial) fault is capble of damping a shear wave 
significantly within one wave cycle.

\section{Discussion and Comparison with Other Theoretical Approaches}\label{s:nine}

We have described a quantitative approach to magnetar activity whose physical
ingredients are i) departures from magnetohydrostatic equilibrium in the core, combined
with ii) the non-linear and temperature-dependent elastic and plastic response of the crust. 
A global treatment of the system is essential:  without it, one cannot begin to understand
the basic timescales, energy scales, and spatial distribution of magnetar activity.  

This framework allows us to make constrained statements about properties as diverse as
the rise times and durations of super-Eddington X-ray bursts; the energies of giant flares
and waiting times between them; the origin of QPO activity and its connection with the
triggering mechanism of SGR bursts; the mechanism by which magnetic energy in the magnetar
crust and core is transmitted to the magnetosphere;  the degree of melting and heating of the
magnetar crust during a burst; the surface covering fraction and pattern of the dissipation;
repeat burst activity; delayed energy release that is manifested as afterglow emission with
a power-law time dependence; and the distribution of electric currents that power the persistent
non-thermal X-ray emission of magnetars.  We are also able to shed some light on why 
super-Eddington bursts are observed only rarely in all magnetars, and at an insignificant level
in many of them, even though all magnetars appear to be strong non-thermal sources with
active magnetospheres and strong torque noise.  

The yielding crust of a magnetar reveals itself to be a very non-linear system in the 
sense that relatively slow variations in one location can trigger much faster variations
in another.  Variability emerges over a wide range of timescales, which turns out to
be computationally challenging in the case where runaway creep is limited by the stretching
of an embedded magnetic field.

We close by comparing our results and conclusions with other theoretical approaches, and
outline some open problems.

\subsection{Non-local Tiggering of Rapid Creep}
Stress build-up and release in the magnetar crust is a {\it non-local} phenomenon,
in good part due to the small aspect ratio of crustal thickness to stellar radius.
A yielding criterion involving a local balance between
Maxwell stress and solid stress \citep{TD1996,perna11,bl14,lander15} does not offer an adequate
description of runaway creep.  On the observational side,
spectral analysis suggests that some slower transient behavior seen in the AXPs involves
{\it global response} of the crust to local dissipation \citep{woods04}.
The non-local yielding process described here also offers a framework for investigating 
small precursor events that have been detected shortly before giant magnetar flares
\citep{hurley05}.  

\subsection{Overstability of Global Elastic Modes}
A consequence of 1. is that a standing elastic wave in the crust will modulate
the creep rate at discrete plastic spots within the crust.  The elastic
wave becomes {\it overstable} if the covering fraction of the plastic spots is large
enough, and if the crust is hydromagnetically unstable within these zones.  
This provides a promising explanation
for the persistence of discrete quasi-periodic oscillations in magnetar flare lightcurves
\citep{israel05,sw05,ws06}.  This mechanism also removes the objection of 
\cite{Levin06} to identifying magnetar QPOs with crustal elastic modes:  we find
that growth can be faster than the damping caused by coupling to a continuous spectrum 
of modes in the magnetized core.  

Magnetar QPOs are therefore identified {\it with periodically forced yielding in 
concentrated shear zones.}   This allows the emission zone of the modulated X-rays
to be fairly compact, smaller than or comparable to $R_\star$.
Indeed the strong rotational modulation detected in the QPO power \citep{sw06}
implies an occultation of the emission zone by the star and {\it requires} a compact size. 
A persistent oscillation of the magnetar crust does also couple to extended dipolar magnetic
field lines outside the star \citep{Timokhin2008}, but only relatively weakly
\citep{TD2001}, and in a way that should be continuously visible over a full
rotation.

\subsection{Intermittency of Super-Eddington Outbursts}
We have uncovered a plausible reason why super-Eddington outbursts are relatively rare
events in magnetars (in comparison with slower but still dramatic changes in persistent X-ray
output and spindown torque, both of which require strong magnetospheric currents).  
The extreme sensitivity of creep rate to applied stress and 
temperature can be compensated by the build-up of a reverse Maxwell stress during plastic flow.  
In a 2-dimensional simulation, the seed magnetic field must be tangled (effectively isotropic)
and exceed a substantial minimum strength.  This does not suppress the formation of 
narrow zones of plastic creep, but it can help to cap the rate of creep within these zones.
As the strength of this embedded field is increased, one sees a transition from a giant-flare
like phenonemon to a slower and less energetic outburst more typical of the transient magnetars.
Super-Eddington outbursts may depend on a favorable magnetic geometry.  

\subsection{Core Hydromagnetic Instability}
A transition from hydromagnetic stability to instability in the core is needed
to explain the most energetic magnetar X-ray flares ($> 10^{45} $ erg).  A significant
structural change in the core magnetic field must come with longer-term side effects:  indeed
our simulations produce repeat bursts of lower energy.   It
should be taken seriously as the underlying driver of lower-energy SGR bursts as well.
We show that such a transition can, even in the absence of a crustal response, produce transient effects
over timescales of a day, significantly shorter than the thermal conduction time across the crust.

\subsection{Characteristic Burst Timescale}
The $\sim 0.1$ s duration seen in a wide range of magnetar X-ray bursts (e.g. \citealt{Gogus2001})
is a signature of stress redistribution within the crust.  We demonstrate
this for large ($\sim 10^{45}$ erg) energy releases
using a global 2-dimensional crust model that includes magnetic, elastic, plastic,
and thermal effects.  In major outbursts, the peak of plastic dissipation jumps by
kilometers, often repeatedly.  Models which postulate that most of the available
energy is stored in a twisted magnetosphere  at the onset of a burst \citep{Lyutikov03} 
have not yet offered a quantitative explanation for the $\sim 0.1$ s timescale.

\null
\vskip.2in

\subsection{Rapid (Millisecond) Dissipation Growth}
Millisecond growth times in the dissipation rate arise from the non-linear and 
temperature dependent plastic response of the crust:  they correspond to stress 
rebalancing over a distance $\sim \delta R_c \sim 0.3$ km.   Rapid horizontal
shear motions in the crust are accompanied by strong vertical magnetospheric currents,
and plausibly drive secondary current-driven instabilities \citep{TD2001,elenbaas16}.  
The very shortest (sub-millisecond) growth seen in giant flares may point to such a
magnetospheric phenomenon \citep{TD1995, Lyutikov03}.

\subsection{Outbursts from Ejection of Crustal Magnetic Shear}
Horizontal shuffling of poloidal magnetic field lines in the core is not
fully mirrored by shuffling in the magnetosphere, and therefore leaves
behind vertical magnetic shear in the crust.   We show that such embedded shear
is rapidly (over several milliseconds) ejected from the crust when the non-local
component of the stress builds up to a critical value.   This is in line with early
theoretical ideas suggesting the ejection of magnetic shear as the source
of low-energy magnetar bursts \citep{TD1995}.  The efficiency of this process
is significantly enhanced if the background magnetic field is already sheared outside
the neutron star.  It will be further enhanced by secondary current-driven instabilities.
The coupling of an elastic wave to the magnetosphere is slower by comparison 
\citep{Blaes1989,TD2001,Link14}.

\subsection{Faults}
Narrow fault-like structures, with a thickness $\sim \delta R_c$,
easily form in the magnetar crust in response to unbalanced
Maxwell stresses at the lower crust boundary.  They provide a promising
explanation for the spectral inference of hotspots on the surfaces
of active magnetars \citep{ibrahim01, lenters03, woods04, esposito07}.  
Faults are especially prominent when the applied Maxwell stress has a rough rotational symmetry
(one example being a kinked toroidal field in the outer core).  The core magnetic field
can otherwise be fairly smooth, and need not contain current sheets.

In contrast with the local analysis
of \cite{LL12}, we find that faults need not be suppressed when the embedded crustal
magnetic field lies transverse to the preferred direction of yielding.
That is because plastic creep with a narrow fault is driven by a stress 
accumulated over a much wider area of crust, so that magnetic shearing 
within the fault remains energetically feasible.  

\subsection{Melting}
Our 2-dimensional model shows that the crust is nearly fully melted within
active faults during major outbursts.  We show that a standing Alfv\'en wave
in a vertically magnetized liquid crust damps relatively slowly, and oscillates
with  a $\sim$ kHz frequency.  Melted faults are promising locations for
the high-frequency QPO activity seen in giant magnetar flares \citep{sw06}.

\null
\vskip .4in

\subsection{X-ray Afterglow from Continuing Crustal Creep}
The same 2-dimensional model shows extended plastic dissipation in the aftermath 
of a giant energy release, declining as a power-law $\sim t^{-1}$ and releasing
a net energy $\sim 10^{-2}$ of the outburst energy.  These properties are remarkably
similar to the X-ray afterglow profile detected following the
1998 giant flare of SGR 1900$+$14 \citep{woods01}.  

One extracts from this example a broader lesson that 
may apply to other instances of transient X-ray emission (e.g.
to the slower outbursts of transient magnetars:  \citealt{ibrahim04,mori13,ah16}, and references therein).  
The magnetospheric twist associated with transient emission need not be injected all at once,
as in the model of \cite{beloborodov09}, and is probably localized closer to the star
than is implied by the dipolar `j-bundle' construction.

\subsection{Some Open Questions}

{\it Implications for magnetar eigenvalue problem.}  The potential of magnetar QPOs as a diagnostic 
of the internal structure has inspired detailed approaches to the eigenvalue problem by
several groups \citep{piro05,glampedakis06,samuelsson07,sotani08,cerda09,colaiuda11,gabler11,vanhoven11,
vanhoven12,passamonti16}.  The eigensolutions are not only sensitive to magnetic
and compositional effects in the elastic parts of the crust, but are also subtly modified by plastic flow.
In particular, creep in localized faults may help generate the observed QPO frequency drifts by introducing
phase shifts into shear wave propagation.  

{\it Hard and persistent X-ray emission.}   Magnetars are copious sources of hard X-rays
in quiescence, with a power greatly exceeding the spindown power \citep{murakami94,kouveliotou99,kuiper06}.
This emission
is probably a signature of very strong electric currents flowing through the closed magnetosphere.
The source location has variously been ascribed to compact
and mildly relativistic plasma near the magnetar surface \citep{TB05}; or to outward
relativistic flows of pairs along a thin bundle of closed field lines surrounding 
the magnetic dipole axis \citep{beloborodov13}.  Our 2-dimensional simulations naturally produce compact
currents associated with strong gradients in creep rate, and distributed broadly across the crust.  
They therefore
favor the first, more localized, type of X-ray emission process.  The pulsed radio emission detected
from some transient magnetars \citep{camilo06,camilo07} also gives evidence for a dynamic magnetosphere;
but it transmits much less energy than the X-rays and is much more rapidly variable, and 
is consistent with emission from open (or nearly open) magnetic field lines \citep{T08}.

{\it Thermalization of magnetic shear energy ejected from magnetar crusts.}
We do not address how vertical or horizontal magnetic shear will damp after
injection into the magnetosphere.    Emission
of quasi-thermal X-rays during super-Eddington magnetar bursts depends on
a transfer of energy from relatively large scale ($\gtrsim \delta R_c$) 
magnetic gradients to sub-relativistic particles with gyrational radii
some $10^{-17}$ times smaller.  Some type of cascade process must be involved
\citep{TB1998}.

Plastic dissipation in the upper crust is found to be negligible, in agreement
with \cite{li2015}.  We nonetheless note that any cascade process that
operates outside the star should also operate {\it within} the upper crust
(where $B^2/8\pi \gg \rho c^2$ and $V_A \sim c$).  The profile of heat deposition may therefore be flat
compared with the pressure profile \citep{lyubarsky02}.  
Prompt burst afterglow emission is a probe of the thermalization process.

{\it Surface Maxwell stresses.}  Persistent electric currents flowing
through the surfaces of magnetars would cut the
vertical imbalance in $B_\perp B_z/4\pi$.  The outer core and inner crust 
generally can sustain a stronger magnetic twist than the magnetosphere, but 
this effect could be important in zones with modest internal twist. 

{\it Do lower-energy SGR bursts represent a `bottom-up' or `top-down' phenomenon?}
Earthquakes are ultimately the consequence of large-scale convective motions and
compositional divisions within the Earth \citep{turcotte02}.  
We have shown that fault-like structures emerge naturally in magnetar crusts,
but our global model does not resolve features smaller than $\sim 0.3$ km.  
Therefore we cannot offer a quantitative model for the origin of low-energy SGR bursts.

The $\sim 0.1$ s characteristic duration of low-energy SGR bursts
does highlight an important difference with earthquakes:  this duration
is comparable to the shear-wave propagation time across the magnetar crust, 
pointing to strongly non-local energy release;  whereas the terrestrial events are
relatively short by the analogous measure.  
In spite of this distinction, the energy distributions of SGR X-ray bursts and earthquakes
follow similar power laws, with smaller bursts supplying a smaller cumulative 
energy release than the largest flares \citep{cheng96,Gogus2001}.  

Finally, we note that a role for a large-scale hydromagnetic instability in facilitating
low-energy burst activity is suggested by the relative activity of the same sources which
(so far) have produced giant flares.   

{\it What role does Hall drift play in driving transient magnetar phenomena?}
Our approach here is partly motivated by the fact that the
rate of crustal Hall drift may be exceeded in young and hot magnetars by the rate
by magnetic drift in the core \citep{Haensel90,GR92}, given the strong feedback of magnetic dissipation
on the temperature and therefore on the ability of the magnetic field to overcome the backpressure
induced by the radial electron-fraction gradient \citep{TD1996}.  

The sign of the effect of Hall drift on magnetar activity is not clear.  
The time dependent yielding calculations presented here raise the possibility that Hall drift 
has a {\it calming} effect, opposite to one usually considered.
By sourcing small-scale magnetic irregularities \citep{GR92,gourg16}, it helps to create
components of the magnetic field transverse to the local direction of shear,
which we find are key to suppressing runaway creep.  Alternatively, small zones of 
enhanced magnetic flux density driven by Hall drift could become preferred dissipation sites 
in the presence of global stresses.  The Lorentz force density increases with wavenumber
in an electron-MHD cascade \citep{cho04}.

The relevance of Hall drift for {\it large-scale} reorganizations of the magnetic field depends on the 
the location of the currents that support the stellar magnetic field.   Large-scale simulations
\citep{perna11,gourg16}
achieve a large redistribution of the dipole magnetic flux within the active lifetimes of a magnetar 
($\sim 10^3$-$10^4$ yr) only if the external field is anchored in the crust, so that the crustal toroidal field 
approaches $10^{16}$ G.  Otherwise the latitudinal electron drift is too slow to transport the poloidal 
magnetic field a significant distance.

\acknowledgments
HY thanks Jonah Miller for discussion of the numerical implementation of wave equations.
This research was supported in part by NSERC, and in part by the Perimeter Institute for
Theoretical Physics.  Research at Perimeter Institute is supported by the Government of
Canada through the Department of Innovation, Science and Economic Development Canada, 
and by the Province of Ontario through the Ministry of Research and Innovation. 

\begin{appendix}

\section{Localized Plastic Spot}

Here we give a concrete example of a plastically deforming circular spot in a planar shell
that is subject to a Maxwell stress $B_\phi B_z/4\pi$ at its lower boundary.  The horizontal
gradient scale of the applied stress is large everywhere compared with the shell thickness
$\delta R_c$, so that magnetoelastic equilibrium can be described in terms of a planar stress
$\sigma_{\phi\varpi} = \bar\mu\varpi \partial_\varpi(\xi_\phi/\varpi)$.  The toroidal 
displacement field is assumed to be independent of height, and $\bar\mu$ is the vertically
mass-averaged shear modulus.  

The vertical magnetic field $B_z$ is everywhere uniform, and we take
\be
B_\phi = B_0 { (\varpi/r_0)^3 \over (1 + \varpi^2/r_0^2)^4},
\ee
so that the Maxwell stress is strongly localized at $\varpi \lesssim r_0$.
The solution to the equation of magneto-elastic equilibrium
\be
{\delta R_c \over\varpi^2}{\partial\over\partial\varpi}\left(\varpi^2\sigma_{\phi\varpi}\right)
= {B_\phi B_z\over 4\pi}\biggr|_-
\ee
is 
\be
\sigma_{\phi\varpi} = {B_0B_zr_0\over 24\pi \bar\mu \delta R_c} \left({\varpi^2/r_0^2\over 1 + \varpi^2/r_0^2}\right)^3.
\ee
This peaks at $\varpi = r_p = \sqrt{2} r_0$.   The displacement field
\be
{\xi_\phi\over r_0} = {B_0B_zr_0\over 48\pi \bar\mu \delta R_c} 
          {(\varpi/r_0)(\varpi^2/r_0^2 + 0.5)\over (1 + \varpi^2/r_0^2)^2}
\ee
scales as $\varpi^{-1}$ at large radius.

Time-dependence of the applied stress corresponds to $\partial_t B_0 \neq 0$.  The net circulation is given by
\be
{\Gamma\over 2\pi} = \varpi \dot\xi_\phi =  {27\over 16} r_p^2 {\partial_t\sigma_{\phi\varpi}|_{\rm max}\over \bar\mu}.
\ee

\end{appendix}


\begin{thebibliography}{}
\bibitem[Akiyama et al.(2003)]{AW03} Akiyama, S., Wheeler, J.~C., Meier, D.~L., 
\& Lichtenstadt, I.\ 2003, \apj, 584, 954 
\bibitem[Alford \& Halpern(2016)]{ah16} Alford, J.~A.~J., \& Halpern, J.~P.\ 2016, \apj, 818, 122 
\bibitem[Arras et al.(2004)]{arras04} Arras, P., Cumming, A., \& Thompson, C.\ 2004, \apjl, 608, L49 
\bibitem[Beloborodov(2009)]{beloborodov09} Beloborodov, A.~M.\ 2009, \apj, 703, 1044 
\bibitem[Beloborodov(2013)]{beloborodov13} Beloborodov, A.~M.\ 2013, \apj, 762, 13 
\bibitem[Beloborodov \& Thompson(2007)]{BT07} Beloborodov, A.~M., \& Thompson, C.\ 2007, \apj, 657, 967 
\bibitem[Beloborodov \& Levin(2014)]{bl14} Beloborodov, A., \& Levin, Y. 2014, \apjl, 794, L24
\bibitem[Blaes et al.(1989)]{Blaes1989} Blaes, O., {\em et al}. 1989, \apj, 343, 839
\bibitem[Braithwaite(2008)]{braithwaite08} Braithwaite, J.\ 2008, \mnras, 386, 1947 
\bibitem[Camilo et al.(2006)]{camilo06} Camilo, F., Ransom, S.~M., Halpern, J.~P., et al.\ 2006, \nat, 442, 892 
\bibitem[Camilo et al.(2007)]{camilo07} Camilo, F., Ransom, S.~M., Halpern, J.~P., \& Reynolds, J.\ 2007, \apjl, 666, L93 
\bibitem[Cerd{\'a}-Dur{\'a}n et al.(2009)]{cerda09} Cerd{\'a}-Dur{\'a}n, P., Stergioulas, N., \& Font, J.~A.\ 2009, \mnras, 397, 1607 
\bibitem[Chabrier(1993)]{chabrier93} Chabrier, G.\ 1993, \apj, 414, 695 
\bibitem[Cheng et al.(1996)]{cheng96} Cheng, B., Epstein, R.~I., Guyer, R.~A., \& Young, A.~C.\ 1996, \nat, 382, 518 
\bibitem[Cho \& Lazarian(2004)]{cho04} Cho, J., \& Lazarian, A.\ 2004, \apjl, 615, L41 
\bibitem[Chugunov \& Horowitz(2010)]{Chugunov2010} Chugunov, A. I., \& Horowitz, C. J. 2010, \mnras, 407, L54
\bibitem[Colaiuda \& Kokkotas(2011)]{colaiuda11} Colaiuda, A., \& Kokkotas, K.~D.\ 2011, \mnras, 414, 3014 
\bibitem[Elenbaas et al.(2016)]{elenbaas16} Elenbaas, C., Watts, A.~L., Turolla, R., \& Heyl, J.~S.\ 2016, \mnras, 456, 3282 
\bibitem[Esposito et al.(2007)]{esposito07} Esposito, P., Mereghetti, S., Tiengo, A., et al.\ 2007, \aap, 461, 605 
\bibitem[Gabler et al.(2011)]{gabler11} Gabler, M., Cerd{\'a} Dur{\'a}n, P., Font, J.~A., M{\"u}ller, E., \& Stergioulas, N.\ 2011, \mnras, 410, L37 
\bibitem[Gill \& Heyl(2010)]{gill10} Gill, R., \& Heyl, J.~S.\ 2010, \mnras, 407, 1926 
\bibitem[Glampedakis et al.(2006)]{glampedakis06} Glampedakis, K., Samuelsson, L., \& Andersson, N.\ 2006, \mnras, 371, L74 
\bibitem[Goldreich \& Reisenegger(1992)]{GR92} Goldreich, P., \& Reisenegger, A. 1992 \apj, 395, 250
\bibitem[G{\"o}{\v g}{\"u}{\c s} et al.(2000)]{Gogus2000} G{\"o}{\v g}{\"u}{\c s}, E., {\em et al}. 2000, \apj, 532, L121
\bibitem[G{\"o}{\v g}{\"u}{\c s} et al.(2001)]{Gogus2001} G{\"o}{\v g}{\"u}{\c s}, E., Kouveliotou, C., Woods,P.~M.,
et al.\ 2001, \apj, 558, 228 
\bibitem[Gourgouliatos et al.(2016)]{gourg16} Gourgouliatos, K., Wood T., and Hollerbach, R., Proc. Nat. Acad. Sci.
(arXiv 1604.01399)
\bibitem[Haensel et al.(1989)]{Haensel1989} Haensel, P., \& Zdunik, J.. L., \& Dobaczewski, J. 1989, \aap, 222, 353
\bibitem[Haensel et al.(1990)]{Haensel90} Haensel, P., Urpin, V.~A., \& Iakovlev, D.~G.\ 1990, \aap, 229, 133 
\bibitem[Hill(1998)]{hill98}  Hill, R. 1998, The Mathematical Theory of Plasticity (Oxford: OUP)
\bibitem[Ho et al.(2015)]{Ho15} Ho, W.~C.~G., Elshamouty, K.~G., Heinke, C.~O., \& Potekhin, A.~Y.\ 2015, \prc, 91, 015806 
\bibitem[Hoffman \& Heyl(2012)]{hoffman12} Hoffman, K., \& Heyl, J.\ 2012, \mnras, 426, 2404 
\bibitem[Hurley et al.(2005)]{hurley05} Hurley, K., Boggs, S.~E., Smith, D.~M., et al.\ 2005, \nat, 434, 1098 
\bibitem[Ibrahim et al.(2001)]{ibrahim01} Ibrahim, A.~I., Strohmayer, T.~E., Woods, P.~M., et al.\ 2001, \apj, 558, 237 
\bibitem[Ibrahim et al.(2004)]{ibrahim04} Ibrahim, A.~I., Markwardt, C.~B., Swank, J.~H., et al.\ 2004, \apjl, 609, L21 
\bibitem[Israel et al.(2005)]{israel05} Israel, G.~L., Belloni, T., Stella, L., et al.\ 2005, \apjl, 628, L53 
\bibitem[Jackson(1999)]{Jackson1999} Jackson, J. D. 1999 {\it Classical Electrodynamics}, Third edition, Chapter 10
\bibitem[Jones(2003)]{jones03} Jones, P.~B.\ 2003, \apj, 595, 342 
\bibitem[Kouveliotou et al.(1999)]{kouveliotou99} Kouveliotou, C., Strohmayer, T., Hurley, K., et al.\ 1999, \apjl, 510, L115 
\bibitem[Kuiper et al.(2006)]{kuiper06} Kuiper, L., Hermsen, W., den Hartog, P.~R., \& Collmar, W.\ 2006, \apj, 645, 556 
\bibitem[Lander et al.(2015)]{lander15} Lander, S.~K., Andersson, N., Antonopoulou, D., \& Watts, A.~L.\ 2015, \mnras, 
449, 2047 
\bibitem[Lattimer \& Mazurek(1981)]{Lattimer1981} Lattimer, J. M., \& Mazurek, T. J. 1981, \apj, 246 955
\bibitem[Lenters et al.(2003)]{lenters03} Lenters, G.~T., Woods, P.~M., Goupell, J.~E., et al.\ 2003, \apj, 587, 761 
\bibitem[Levin(2006)]{Levin06} Levin, Y. 2006, \mnras, 368, L35
\bibitem[Levin(2007)]{Levin07} Levin, Y.\ 2007, \mnras, 377, 159 
\bibitem[Levin \& Lyutikov(2012)]{LL12} Levin, Y., \& Lyutikov, M.\ 2012, \mnras, 427, 1574 
\bibitem[Li \& Beloborodov(2015)]{li2015} Li, X., \& Beloborodov, A.~M.\ 2015, \apj, 815, 25
\bibitem[Li et al.(2016)]{li16} Li, X., Levin, Y., \& Beloborodov, A.~M.\ 2016, arXiv:1606.04895 
\bibitem[Link(2014)]{Link14} Link, B.\ 2014, \mnras, 441, 2676 
\bibitem[Lyubarsky et al.(2002)]{lyubarsky02} Lyubarsky, Y., Eichler, D., \& Thompson, C.\ 2002, \apjl, 580, L69 
\bibitem[Lyutikov(2003)]{Lyutikov03} Lyutikov, M.\ 2003, \mnras, 346, 540 
\bibitem[McDermott et al.(1988)]{McDermott1988} McDermott, P. N., \& Van Horn, H. M., \& Hansen, C. J. 1988, \apj, 325, 725
\bibitem[Mereghetti et al.(2015)]{Mereghetti15} Mereghetti, S., Pons, J.~A., \& Melatos, A.\ 2015, \ssr, 191, 315 
\bibitem[Mori et al.(2013)]{mori13} Mori, K., Gotthelf, E.~V., Zhang, S., et al.\ 2013, \apjl, 770, L23 
\bibitem[Murakami et al.(1994)]{murakami94} Murakami, T., Tanaka, Y., Kulkarni, S.~R., et al.\ 1994, \nat, 368, 127 
\bibitem[Negele \& Vautherin(1973)]{Negele1973} Negele, J. W., \& Vautherin, D. 1973, Nuclear Physics, A207, 298
\bibitem[Onsi et al.(2008)]{onsi08} Onsi, M., Dutta, A.~K., Chatri, H., et al.\ 2008, \prc, 77, 065805 
\bibitem[Pacini \& Ruderman(1974)]{Ruderman1974} Pacini, F., \& Ruderman, M. 1974, \nat, 251, 399
\bibitem[Palmer et al.(2005)]{palmer05} Palmer, D.~M., Barthelmy, S., Gehrels, N., et al.\ 2005, \nat, 434, 1107 
\bibitem[Passamonti \& Pons(2016)]{passamonti16} Passamonti, A., \& Pons, J.~A.\ 2016, arXiv:1606.02132 
\bibitem[Perna \& Pons(2011)]{perna11} Perna, R., \& Pons, J.~A.\ 2011, \apjl, 727, L51 
\bibitem[Pethick(1992)]{Pethick92} Pethick, C.~J.\ 1992, Structure and Evolution of Neutron Stars, 115 
\bibitem[Piro(2005)]{piro05} Piro, A.~L.\ 2005, \apjl, 634, L153 
\bibitem[Potekhin et al.(2015)]{potekhin15} Potekhin, A.~Y., Pons, J.~A., \& Page, D.\ 2015, \ssr, 191, 239 
\bibitem[Ruderman(1991)]{ruderman91} Ruderman, M.\ 1991, \apj, 382, 587 
\bibitem[Ruderman et al.(1998)]{ruderman98} Ruderman, M., Zhu, T., \& Chen, K.\ 1998, \apj, 492, 267 
\bibitem[Samuelsson \& Andersson(2007)]{samuelsson07} Samuelsson, L., \& Andersson, N.\ 2007, \mnras, 374, 256 
\bibitem[Schubert \& Yuen(1978)]{schubert78} Schubert, G, \& Yuen, D.A.\ 1978, Tectonophysics, 50, 197
\bibitem[Sotani et al.(2008)]{sotani08} Sotani, H., Kokkotas, K.~D., \& Stergioulas, N.\ 2008, \mnras, 385, L5 
\bibitem[Sotani(2011)]{sotani11} Sotani, H.\ 2011, \mnras, 417, L70 
\bibitem[Stringfellow et al.(1990)]{stringfellow90} Stringfellow, G.~S., Dewitt, H.~E., \& Slattery, W.~L.\ 1990, \pra, 41, 1105 
\bibitem[Strohmayer et al.(1991)]{Strohmayer1991} Strohmayer, T., \& Ogata, S., \& Iyetomi, H., \& Ichimaru, S., 
         \& Van Horn,H. M. 1991, \apj, 375, 679
\bibitem[Strohmayer \& Watts(2005)]{sw05} Strohmayer, T.~E., \& Watts, A.~L.\ 2005, \apjl, 632, L111 
\bibitem[Strohmayer \& Watts(2006)]{sw06} Strohmayer, T.~E., \& Watts, A.~L.\ 2006, \apj, 653, 593 
\bibitem[Tanaka et al.(2007)]{tanaka07} Tanaka, Y.~T., Terasawa, T., Kawai, N., et al.\ 2007, \apjl, 665, L55 
\bibitem[Terasawa et al.(2005)]{terasawa05} Terasawa, T., Tanaka, Y.~T., Takei, Y., et al.\ 2005, \nat, 434, 1110 
\bibitem[Thompson(2008)]{T08} Thompson, C.\ 2008, \apj, 688, 1258-1281 
\bibitem[Thompson \& Duncan(1993)]{TD1993} Thompson, C., \& Duncan, R. 1993, \apj, 408, 194
\bibitem[Thompson \& Duncan(1995)]{TD1995} Thompson, C., \& Duncan, R. 1995, \mnras, 275, 255
\bibitem[Thompson \& Duncan(1996)]{TD1996} Thompson, C., \& Duncan, R. 1996 \apj, 473, 322
\bibitem[Thompson \& Blaes(1998)]{TB1998} Thompson, C., \& Blaes, O.\ 1998, \prd, 57, 3219 
\bibitem[Thompson \& Murray(2001)]{TM2001} Thompson, C., \& Murray, N. 2001, \apj, 560, 339
\bibitem[Thompson \& Duncan(2001)]{TD2001} Thompson, C., \& Duncan, R. 2001, \apj, 561, 980
\bibitem[Thompson et al.(2002)]{TLK} Thompson, C., Lyutikov, M., \& Kulkarni, S.~R.\ 2002, \apj, 574, 332 
\bibitem[Thompson \& Beloborodov(2005)]{TB05} Thompson, C., \& Beloborodov, A.~M.\ 2005, \apj, 634, 565 
\bibitem[Timokhin et al.(2008)]{Timokhin2008} Timokhin, A. N., \& Eichler, D., \& Lyubarsky, Y. 2008, \apj, 680, 1398
\bibitem[Turcotte \& Schubert(2002)]{turcotte02} Turcotte, D.L., \& Schubert G. 2002, Geodynamics (Cambridge: CUP)
\bibitem[Turolla et al.(2015)]{Turolla15} Turolla, R., Zane, S., \& Watts, A.~L.\ 2015, Reports on Progress in Physics, 78, 116901 
\bibitem[van Hoven \& Levin(2011)]{vanhoven11} van Hoven, M., \& Levin, Y.\ 2011, \mnras, 410, 1036 
\bibitem[van Hoven \& Levin(2012)]{vanhoven12} van Hoven, M., \& Levin, Y.\ 2012, \mnras, 420, 3035 
\bibitem[Watts \& Strohmayer(2006)]{ws06} Watts, A.~L., \& Strohmayer, T.~E.\ 2006, \apjl, 637, L117 
\bibitem[Woods et al.(2001)]{woods01} Woods, P.~M., Kouveliotou, C., G{\"o}{\v g}{\"u}{\c s}, E., et al.\ 2001, \apj, 552, 748 
\bibitem[Woods et al.(2004)]{woods04} Woods, P.~M., Kaspi, V.~M., Thompson, C., et al.\ 2004, \apj, 605, 378 
\bibitem[Woods \& Thompson(2006)]{WT06} Woods, P. M., \& Thompson, C. 2006, 
{\rm Compact stellar X-ray sources}, Edited by Walter Levin \& Michiel van der Klis. Cambridge University Press
\bibitem[Yakovlev et al.(2001)]{yakovlev01} Yakovlev, D.~G., Kaminker, A.~D., Gnedin, O.~Y., \& Haensel, P.\ 2001, \physrep, 354, 1 



\end{thebibliography}
\end{document}